\documentclass[aps, prx, twocolumn, longbibliography, superscriptaddress, nofootinbib, 10pt]{revtex4-2}

\usepackage{graphicx}
\usepackage{dcolumn}
\usepackage[utf8]{inputenc}
\usepackage[english]{babel}
\usepackage[T1]{fontenc}
\usepackage{xspace}
\usepackage[dvipsnames]{xcolor}
\usepackage{mathtools, bbm, bm, bbold, amssymb, amsthm, physics, amsfonts, mathrsfs}
\usepackage[version=3]{mhchem}  
\usepackage{array, hhline, tabularx, makecell}
\usepackage{dsfont, pifont, yfonts, soul, textcomp} 
\usepackage{enumitem}
\usepackage{pstricks}
\usepackage{multirow}
\usepackage{verbatim}
\usepackage{empheq}
\usepackage[normalem]{ulem}
\definecolor{color1bg}{HTML}{b890a1}
\usepackage{eurosym}
\usepackage{siunitx}
\usepackage{booktabs}
\usepackage[babel]{csquotes}
\usepackage{listings}
\usepackage{hyperref}
\hypersetup{
 colorlinks,
 citecolor=color1bg,
 filecolor=black,
 linkcolor=color1bg,
 urlcolor=black}

\let\tr\relax
\DeclareMathOperator{\tr}{Tr}

\DeclareMathOperator*{\argmin}{argmin}
\DeclareMathOperator*{\argmax}{argmax}
\let\var\relax
\DeclareMathOperator{\var}{Var}
\DeclareMathOperator{\prob}{Prob}

\allowdisplaybreaks

\newcommand{\id}{\mathbbm{1}}
\newcommand{\tpm}{\mathrm{TPM}}
\newcommand{\vom}{\text{VOM}\xspace}
\newcommand{\M}{\mathcal{M}}
\newcommand{\U}{\mathcal{U}}
\newcommand{\J}{\mathcal{J}}
\newcommand{\D}{\mathcal{D}}
\newcommand{\F}{\mathcal{F}}
\newcommand{\av}[1]{\langle #1 \rangle}

\definecolor{green1}{rgb}{0.33, 0.7, 0.69}

\begin{document}

\title{Minimally invasive measurement of work in coherent quantum systems}

\author{Cyril Elouard}
\thanks{cyril.elouard@univ-lorraine.fr}
\affiliation{Universit\'e de Lorraine, CNRS, LPCT, F-54000 Nancy, France}

\author{Karen Hovhannisyan}
\thanks{karen.hovhannisyan@uni-potsdam.de}
\affiliation{University of Potsdam, Institute of Physics and Astronomy, Karl-Liebknecht-Str.~24-25, 14476 Potsdam, Germany}

\author{Giulia Rubino}
\thanks{giulia.rubino@bristol.ac.uk}
\affiliation{H. H. Wills Physics Laboratory, University of Bristol, Tyndall Avenue, Bristol, BS8 1TL, United Kingdom}
\affiliation{Quantum Engineering Technology Labs, H. H. Wills Physics Laboratory and School of Electrical, Electronic, and Mechanical Engineering, University of Bristol, BS8 1FD, UK}

\date{\today}

\begin{abstract}
A central challenge in quantum thermodynamics is to access work fluctuations in coherent processes without distorting the energetics of the unmeasured evolution. In standard two-point schemes, the initial energy measurement dephases coherent inputs, causing the measured average work to differ from that of the unmeasured evolution. Here, we develop an operational scheme for accessing work statistics for closed quantum systems based on the abstract notion of variation in the Heisenberg picture Hamiltonian. This scheme preserves energetically relevant coherences, thereby faithfully reproducing unmeasured work, while still producing positive probabilities. We derive modified Jarzynski and Crooks relations, as well as a thermodynamic uncertainty relation, identifying coherence-induced correction terms. Furthermore, we show that this scheme can reliably quantify the performance of a coherent engine in situations where the two-point energy measurement would suppress work output. In addition, the scheme requires only a single measurement and can predict the work associated with a subsequent unitary transformation. We exploit this feature to construct a Maxwell-demon protocol that can outperform energy-based feedback engines for coherent work extraction. Our results establish this scheme as a framework for accessing coherent work fluctuations without erasing the coherence that drives quantum thermodynamic performance.
\end{abstract}

\maketitle

\section{Introduction}

Driven quantum systems can exchange energy in ways that have no classical analogue, with coherence between energy levels acting both as a hallmark of quantum behaviour, and as an operational resource~\cite{Streltsov2017, Lostaglio2015, Kammerlander2016}. As quantum control improves, a central question that emerges for quantum thermodynamics is how to quantify work when coherence is integral to a device's operation. This matters not only for the conceptual formulation of work in coherent quantum processes, but also for the practical assessment of quantum engines, refrigerators, and other driven platforms whose performance may depend on coherent dynamics~\cite{Kosloff2014, Uzdin_2015, Maslennikov2019, Zhang2022, Cangemi2024}.

A broadly used framework for fluctuating work is the two-point measurement (TPM) scheme, in which energy is measured before and after the driving protocol~\cite{Kurchan2000, Tasaki2000, Talkner_2007, Esposito2009, Campisi2011}. This approach has significant merits: it yields a well-defined stochastic work variable and underpins the standard operational formulations of the Jarzynski and Crooks relations in driven quantum systems \cite{Jarzynski1997, Crooks_1999, Piechocinska_2000, Kurchan2000, Tasaki2000, Jarzynski2006, Jarzynski2011}. However, these advantages come at a cost: for initial states carrying coherence in the energy basis, the first projective measurement dephases the system and thereby alters the subsequent evolution. In such cases, the results of the second measurement are so affected by this back-action that the TPM scheme no longer probes the energetics of the original coherent process. Instead, it measures those of a modified protocol in which coherence is removed before the work exchange phase of the process even begins. In other words, it partially erases the quantumness of the process (e.g., suppressing contextuality \cite{Lostaglio2018}).

This issue has sparked extensive debate about how work should be defined in coherent quantum systems~\cite{Allahverdyan_2005, Talkner_2007, Horodecki_2013, Skrzypczyk_2014, Allahverdyan_2014, Solinas2015, Talkner_2016, Perarnau2017, Baumer2018, Lostaglio2018, Pei_2023, Hovhannisyan_2024, PintoSilva_2025}. A particularly relevant line of work has argued that changes of quantum observables during a unitary process can themselves be represented, in the Heisenberg picture, by a single observable, whose statistics are obtained from its spectral measure~\cite{Bochkov_1977, Chernyak_2004, Allahverdyan_2005, PintoSilva2024, PintoSilva_2025}. When the observable is the Hamiltonian, this construction gives rise, for closed thermally isolated processes, to the so-called Heisenberg work operator. In the context of energy changes, measuring this operator provides an alternative to TPM: instead of reconstructing work from two energy outcomes, one directly measures the observable associated with the energy change generated by the coherent unitary dynamics.
This choice is not arbitrary. Among all projective measurements, the spectral measurement of the Heisenberg work operator is selected by a minimal operational requirement: for every input state, the average result of a work measurement used to diagnose a coherent process should equal the average energy change that would occur in the untouched unitary process.
The recent analysis of Ref.~\cite{PintoSilva_2025} took this even further by showing that this observable-based protocol is the only one satisfying a minimal set of general physically motivated consistency requirements for variations of quantum observables. 

Here, we take the uniqueness of this observable-based scheme---which we refer to as the \emph{variation-operator measurement} (\vom) scheme---as a starting point and ask what thermodynamic structure and operational consequences follow from it. 
First, we refine the protocol by optimizing the measurement instruments to minimize back-action. This also makes explicit how the measurement can be implemented.
Then, we explore the consequence of adopting this operational notion of work on fluctuation relations. Since the random variable being measured is no longer the TPM energy difference, the standard fluctuation relations cannot be expected to remain unchanged~\cite{Hovhannisyan_2024, Rubino_2025}. Instead, we derive the fluctuation structure naturally associated with the \vom scheme. We show that the exponential work average obeys a Jarzynski-like relation with a non-negative correction governed by non-commutativity, and we identify a corresponding backward protocol, a Crooks-like irreversibility measure, and a thermodynamic uncertainty bound for odd observables under time reversal. We then apply the scheme to coherent thermodynamic processes. In simple qubit protocols, \vom and TPM agree in incoherent limits but differ once coherence contributes to the energy exchange. In a four-stroke coherent heat engine, \vom captures the leading coherence-assisted contribution to the extracted work, whereas TPM suppresses it through the initial energy measurement. Finally, we show that \vom can also be used in a feedback-based Maxwell-demon protocol, where the measurement outcome directly predicts whether applying a given unitary will be favourable for work extraction. These examples demonstrate that the \vom scheme is not merely a different representation of work statistics: in coherent regimes, it can lead to quantitatively different operational assessments of thermodynamic performance.

The remainder of this paper is organised as follows. In Sec.~\ref{subsec:VOMscheme}, we introduce the work-operator measurement scheme. We then derive its fluctuation structure: a Jarzynski-like relation (Sec.~\ref{subsec:Jarzynski-like_relation}), a Crooks-like irreversibility measure (Sec.~\ref{subsec:Crooks-like_relation}), and a thermodynamic uncertainty bound (Sec.~\ref{subsec:TUR}). In Sec.~\ref{subsec:case_study}, we illustrate the scheme in a driven qubit, where \vom and TPM can be compared analytically in the presence of coherence. In Sec.~\ref{subsec:4stroke_heat_engine}, we apply the framework to a four-stroke coherent heat engine, showing how \vom captures coherence-assisted work extraction that is suppressed by TPM. In Sec.~\ref{subsec:Maxwell_demon}, we develop a feedback-based Maxwell-demon protocol built from \vom outcomes, highlighting both the advantages and costs of using the work operator as a control diagnostic. Finally, we discuss the broader implications of this operational notion of work for coherent quantum thermodynamics.

\vfill


\section{Work-operator measurement scheme}
\label{subsec:VOMscheme}

We begin by formulating a minimal operational requirement for a measurement scheme intended to assign work values to a coherently driven unitary process. Specifically, we consider a closed quantum system initially prepared in a state $\rho$, whose Hamiltonian is driven from $H(t_\mathrm{i})=H$ at the initial time $t_\mathrm{i}$ to $H(t_\mathrm{f})=H'$ at the final time $t_\mathrm{f}$, generating the unitary evolution $U$. We refer to this as the ``untouched'' process. Since the system is closed, no heat is exchanged during the process, and the average work performed on it is equal to its average energy change:
\begin{align}
    \av{W}_\mathrm{unt} = \tr(H' U \rho U^\dagger) - \tr(H \rho) = \tr(\rho \Omega).
\end{align}
Here,
\begin{align}\label{eq:Omega}
    \Omega := U^\dagger H' U - H
\end{align}
is the variation of the Hamiltonian in the Heisenberg picture, also known as the \textit{operator of work}~\cite{Bochkov_1977, Chernyak_2004, Allahverdyan_2005}. Our aim is to measure the energy change in such a way that the resulting statistics reproduces the untouched work for every input state $\rho$. Namely, we require that
\begin{align} \label{requirement}
    \av{W}_\mathrm{meas} = \av{W}_\mathrm{unt}.
\end{align}
Keeping in mind that any combination of evolution and measurement can be described by a positive operator-valued measure (POVM) $\{ M_a\}_a$ and associated real outcomes $\{w_a\}_a$, we can write the most general form of Eq.~\eqref{requirement} as
\begin{align}
    \av{W}_\mathrm{meas} = \sum_a w_a  \tr(\rho M_a) = \tr(\rho \Omega), \quad \forall \rho.
\end{align}
Since this equality must hold for all density operators $\rho$, it follows that the sought measurement scheme must verify
\begin{align} \label{eq:moment_condition}
    \sum_a w_a M_a = \Omega.
\end{align}
At this level, this condition does not uniquely determine the measurement scheme. In general, different POVMs may satisfy Eq.~\eqref{eq:moment_condition} and, even for a fixed POVM, different measurement instruments (namely, different sets of Kraus operators realising the same effects) may produce the same outcome statistics while inducing different post-measurement states.

The straightforward choice is to take the $M_a$ to be the spectral projectors $\Pi_k \coloneqq \dyad{\omega_k}$ of $\Omega$ and to assign the corresponding eigenvalues $\omega_k$ as outcomes $w_a$. This observable-based construction has been termed a work-operator scheme~\cite{Bochkov_1977, Chernyak_2004, Allahverdyan_2005}. Here, we refer to it as the \textit{variation-operator measurement} (\vom) scheme, to emphasise the fact that the same construction applies more generally to this variation of any physical observable under a unitary process, and not only to energy, where the variation is interpreted as work. Importantly, Ref.~\cite{PintoSilva_2025} recently showed that the \vom scheme is uniquely selected by three broad physical principles. Namely, it is the \textit{only} state-independent scheme that simultaneously satisfies an appropriately defined energy-conservation law, reality, and no-signaling. Its fulfilment of Eq.~\eqref{eq:moment_condition} then follows as a direct consequence of the reality condition.

Although the conditions above uniquely select the POVM and outcomes of the \vom scheme, they do not concretize the instruments (channels that realize the state projection) of the measurement. The standard implicit choice is the L\"{u}ders prescription, which posits that the Kraus operators of the instruments are simply $\dyad{\omega_k}$~\cite{Luders_1951}. However, any set of Kraus operators of the form $\ketbra{\psi_k}{\omega_k}$, with arbitrary $\{\ket{\psi_k}\}_{k=1}^d$, yields the same POVM $\{ \Pi_k\}$. Here, we refine the \vom scheme by finding the optimal set of Kraus operators that minimize the measurement back-action. Namely, in Sec.~\ref{SM:back-action} of the Supplementary Material, we prove that the Kraus operators
\begin{align}
    K_k^\Omega = U \Pi_k,
\end{align}
where the superscript $\Omega$ denotes the \vom scheme, simultaneously minimise \textit{i.} the disturbance of the state with respect to the final state of the untouched process and \textit{ii.} the effect of the measurement on the energetics of the process.

In the Schr\"{o}dinger picture, the measurement channel defined by $\{K_k^\Omega\}_{k=1}^d$ effectively consists of two operations: a projection onto $\Pi_k$, followed by the unitary evolution $U$. Since $\Omega$ depends on $U$, this implementation of the \vom scheme is somewhat proleptic: it requires full knowledge of $U$ and $H'$ already at the initial time, before the process takes place.
Each outcome of a measurement of $\Omega$ at $t_\mathrm{i}$ can be understood as the amount of work that \textit{would be performed upon subsequently applying the unitary $U$ to the system}. This work is therefore not measured during the application of the unitary itself. At the same time, access to the work value before applying $U$ opens the possibility of feedback-based protocols, for example by applying the unitary only when the outcome satisfies a desired criterion (see Sec.~\ref{subsec:Maxwell_demon}).
Alternatively, the same minimal back-action can be achieved by applying the Kraus operators $U \Pi_k U^\dagger$ to the final state $U \rho U^\dagger$. Namely, a standard L\"{u}ders projective measurement of $U \Omega U^\dagger$ on $U \rho U^\dagger$ at $t_\mathrm{f}$ also realizes a minimally invasive \vom measurement channel. This implementation of the \vom scheme is more causally intuitive, as it requires no knowledge of future stages of the process.

\medskip

Although $\{K_k^\Omega\}$ specify a well-defined measurement channel, directly engineering a system--apparatus interaction that measures an operator such as $\Omega$ can be nontrivial~\cite{Chernyak_2004, PintoSilva_2025}. A natural operational implementation of the \vom scheme can instead be obtained by introducing a unitary $R$ such that $\ket{\omega_k} = R\ket{E_k}$. The protocol then consists of applying $R^\dagger$, performing a projective energy measurement, and subsequently applying $UR$. Altogether, these steps realise the instrument associated with $K_k^\Omega$.

Lastly, while evaluating work from a single measurement may seem at odds with the common thermodynamic statement that work is a path-dependent quantity~\cite{Talkner_2007}, we argue in Appendix~\ref{app:Work_Obs} that this statement does not apply to the closed (unitary) dynamics considered here. We also discuss in that appendix how this scenario encompasses the open-system case.

\section{Fluctuation relations for the \vom scheme}
\label{sec:fluc_rel}

\subsection{Jarzynski-like fluctuation relation}
\label{subsec:Jarzynski-like_relation}

Having introduced the \vom scheme, we now ask what fluctuation-type relation replaces the standard Jarzynski equality for this operational definition of work. It was shown in Ref.~\cite{Hovhannisyan_2024} that, whenever $[U^\dagger H' U,H]\neq 0$, any scheme satisfying the standard Jarzynski equality for all inverse temperatures $\beta$ cannot simultaneously reproduce the untouched average work for all input states. Thus, once the latter requirement [Eq.~\eqref{eq:moment_condition}] is imposed, the exact Jarzynski equality is generally lost. Instead, for such schemes (including the \vom) the exponential work average satisfies a modified, Jarzynski-like relation with a nonnegative correction term \cite{Rubino_2025}.

In the setting of the fluctuation relation, the system starts in a thermal (Gibbs) state $\tau_\beta = e^{-\beta H} / Z$, where $\beta = (k_\mathrm{B} T)^{-1}$ is the inverse temperature ($k_\mathrm{B}$ being the Boltzmann constant, $T$ the temperature of the system) and $Z = \tr(e^{-\beta H})$ is the partition function. Then, as is well-known, the TPM scheme yields the standard Jarzynski equality $\bigl\langle e^{-\beta W} \bigr\rangle_\tpm = e^{-\beta \Delta F}$~\cite{Jarzynski1997,Kurchan2000,Tasaki2000}, where $\Delta F = - \beta^{-1} \ln \frac{Z'}{Z}$ is the free energy difference. By contrast, in the \vom scheme, the work outcomes are the eigenvalues of $\Omega$, and therefore, \cite{Allahverdyan_2005}
\begin{align} \label{eq:Jarzynski_VOM_start}
\begin{split}
    \bigl\langle e^{-\beta W} \bigr\rangle_{\Omega} &= \tr\big(\tau_\beta e^{-\beta \Omega}\big)
    \\
    &= \frac{1}{Z}\,\tr \! \big[ e^{-\beta H} e^{-\beta (U^\dagger H' U - H)} \big] \geq \frac{Z'}{Z}.
\end{split}
\end{align}
Here, the second line is due to the Golden--Thompson inequality \cite{BhatiaPDM_2007}, which becomes strict whenever $H$ and $U^\dagger H' U$ do not commute \cite{Petz_1988, Petz_1994}.
Recalling that $\frac{Z'}{Z} = e^{-\beta \Delta F}$, it is convenient to rewrite Eq.~\eqref{eq:Jarzynski_VOM_start} in the Jarzynski-like form
\begin{align} \label{eq:corrected_Jarzynski}
    \left\langle e^{-\beta W} \right\rangle_{\Omega} = e^{-\beta \Delta F + \Xi_\Omega},
\end{align}
where the correction term $\Xi_\Omega$ is defined by
\begin{align} \label{eqn:Xi_def}
\Xi_\Omega := \ln \frac{\tr \big[ e^{-\beta H} e^{-\beta (U^\dagger H' U - H)} \big]}{\tr\left(e^{-\beta U^\dagger H' U}\right)} \geq 0.
\end{align}
Note that the application of Jensen's inequality in Eq.~\eqref{eq:corrected_Jarzynski} yields $\av{W}_\Omega \ge \Delta F - \beta^{-1} \Xi_\Omega$ which is looser than the $\av{W}_\Omega \ge \Delta F$ bound guaranteed by Eq.~\eqref{requirement}. For example, in a cyclic process with $H'=H$, one has $\Delta F=0$, while passivity of the Gibbs state still implies $\av{W}_\Omega\ge 0$, irrespective of the value of $\Xi_\Omega$.

The quantity $\Xi_\Omega$ quantifies by how much the usual Jarzynski lower bound on the average work is relaxed in the \vom framework. This relaxation should be understood as a modification of the fluctuation relation associated with the coherent work observable, rather than as an additional thermodynamic potential. The correction arises because the \vom scheme assigns fluctuations through the spectral statistics of $U^\dagger H'U-H$, instead of through pairs of sharp energy outcomes. When $H$ commutes with $U^\dagger H'U$, the two descriptions coincide and the standard Jarzynski equality is recovered.

Away from this commuting limit, the work values measured by \vom still have a direct energetic meaning for the subsequent unitary evolution, but they cannot in general be resolved into differences of simultaneously well-defined initial and final energies. A positive $\Xi_\Omega$ nevertheless indicates that the \vom statistics place greater weight on outcomes with $W<\Delta F$ than in the TPM case. More precisely, using Markov's inequality~\cite{Durrett2019} on the \vom work distribution, we obtain from Eq.~\eqref{eqn:Xi_def} that, for any $\zeta > 0$,
\begin{align}
    \prob\bigl(W-\Delta F\le -\zeta\bigr)\le e^{\Xi_\Omega-\beta \zeta}.
\end{align}
Thus, $\Xi_\Omega$ enlarges the allowed tail of such apparent second-law-violating fluctuations (see also the related discussion in Ref.~\cite{Rubino_2025}). In this sense, $\Xi_\Omega$ provides a quantitative measure of how far the \vom statistics depart from the standard Jarzynski picture in coherent, noncommuting regimes.

\subsection{Crooks-like relation}
\label{subsec:Crooks-like_relation}

We next derive an analogue of Crooks's detailed fluctuation relation \cite{Crooks_1999, Piechocinska_2000, Kurchan2000, Tasaki2000} for the \vom scheme. To that end, we start by defining a backward, time-reversed protocol. We denote by $\tilde{Q}$ the time-reversed counterpart of any quantity $Q$ pertaining to the original (``forward'') process. By definition, the time reversal is an involution, so $\tilde{\tilde{Q}} = Q$. For observables, time reversal is represented by the antiunitary operator $\Theta$ such that $\tilde{Q} = \Theta Q \Theta^\dagger$; for example
\begin{align} \label{revhams}
    \tilde{H} = \Theta H \Theta^\dagger \qquad \mathrm{and} \qquad \tilde{H}' = \Theta H' \Theta^\dagger.
\end{align}
The operator $\Theta$ satisfies $\Theta\, i = -\, i\, \Theta$, $\Theta \Theta^\dagger = \id$, and reverses the sign of observables that are odd under time reversal~\cite{Messiah}.

The forward protocol takes the system from $H$ to $H'$ under the unitary $U$, with initial Gibbs state $\tau_\beta$. The VOM work statistics are then assigned by measuring the forward work operator $\Omega$.
The corresponding backward protocol is defined by preparing the system in the Gibbs state $\tilde{\tau}'_\beta = e^{-\beta \tilde{H}'}/Z'$ and driving it from $\tilde{H}'$ to $\tilde{H}$, generating the time-reversed unitary~\cite{Campisi2011, Gaspard2008, Crooks_2008}
\begin{align} \label{revuni}
    \tilde{U} = \Theta U^\dagger \Theta^\dagger
\end{align}
along the way. The associated backward work operator is then $\tilde{\Omega} := \tilde{U}^\dagger \tilde{H} \tilde{U} - \tilde{H}'$, and, in view of Eq.~\eqref{revuni}, it is related to the forward work operator by
\begin{align} \label{eqn:OmegaB_from_OmegaF}
    \tilde{\Omega} = -\,\Theta\, U\, \Omega\, U^\dagger \Theta^\dagger.
\end{align}
This immediately determines the spectral structure of the backward work operator. Indeed, it is easy to check that $\ket{\tilde{\omega}_k} \coloneqq \Theta U \ket{\omega_k}$ is an eigenvector of $\tilde{\Omega}$ with the eigenvalue
\begin{align} \label{eqn:involution_outcomes}
    \tilde{\omega}_k = -\,\omega_k.
\end{align}
Thus, the backward outcome corresponding to the forward outcome $k$ carries the opposite work value, with the associated backward POVM element
\begin{align} \label{eqn:proj_map}
    \tilde{\Pi}_k \coloneqq \dyad{\tilde{\omega}_k} = \Theta\, U\, \Pi_k\, U^\dagger \Theta^\dagger.
\end{align}
Conveniently, the correspondence in Eq.~\eqref{eqn:involution_outcomes} allows us to treat the forward and backward \vom statistics as distributions on a common ``event space'' labeled by $k$. Introducing the forward and backward \vom probability distributions
\begin{align} \label{probs}
    p_k \coloneqq \tr(\tau_\beta \Pi_k) \qquad \mathrm{and} \qquad \tilde{p}_k \coloneqq \tr(\tilde{\tau}'_\beta \tilde{\Pi}_k),
\end{align}
we can thus define the Crooks-style stochastic entropy production
\begin{align} \label{eqn:sigmaOmega_def}
    \sigma^\Omega_k := \ln\biggl(\frac{p_k}{\tilde{p}_k}\biggr). 
\end{align}
Its forward mean is
\begin{align} \label{eqn:sigmaOmega_mean_KL}
    \langle \sigma^\Omega\rangle = \sum_k p_k\,\ln\biggl(\frac{p_k}{\tilde{p}_k}\biggr) = D_{\mathrm{KL}}(p \, \Vert \, \tilde{p}) \geq 0,
\end{align}
namely, the Kullback--Leibler divergence between the forward and backward \vom outcome distributions.

This quantity can be compared directly with the standard TPM entropy production. In fact, using Eqs.~\eqref{revhams} and~\eqref{eqn:proj_map}, we bring the backward probabilities to the form $\tilde{p}_k = \tr(U^\dagger\tau'_\beta U\,\Pi_k)$, where $\tau'_\beta = e^{-\beta H'}/Z'$. Thus, introducing the $\Omega$-dephasing channel $\mathcal{D}_\Omega[\rho] \coloneqq \sum_k \Pi_k \rho \Pi_k$, we can write
\begin{align} \label{KL-to-QRE}
    \av{\sigma^\Omega} = D_\mathrm{KL}\big(p \, \Vert \, \tilde{p}\big) = S\big( \mathcal{D}_\Omega[\tau_\beta] \, \big\Vert \, \mathcal{D}_\Omega[U^\dagger \tau'_\beta U] \big),
\end{align}
where $S(\cdot \, \Vert \, \cdot)$ is the quantum relative entropy \cite{Nielsen_Chuang_2010}. Applying the data-processing inequality~\cite{Nielsen_Chuang_2010} to the relative entropy in Eq.~\eqref{KL-to-QRE}, we finally arrive at
\begin{align}
    \av{\sigma^\Omega} \leq S(\tau_\beta \, \Vert \, U^\dagger \tau'_\beta U).
\end{align}
Here, the right-hand side is precisely the standard TPM entropy production,
\begin{align*}
    S(\tau_\beta \, \Vert \, U^\dagger \tau'_\beta U) = \beta(\av{W} - \Delta F) = \av{\sigma^\tpm},
\end{align*}
and therefore,
\begin{align} \label{VOM_EP_TPM}
    \av{\sigma^\Omega} \leq \av{\sigma^\tpm}.
\end{align}
This ordering reflects the fact that the \vom scheme retains more information about the probed process than the TPM scheme, thereby generating less entropy.

The departure from the standard Crooks relation can also be seen more explicitly at the level of the forward-to-backward ratio:
\begin{align}
\frac{p_k}{\tilde{p}_k} = e^{\beta (\omega_k - \Delta F)} \frac{\bra{\omega_k} e^{-\beta H} \ket{\omega_k} \bra{\omega_k} e^{-\beta \Omega} \ket{\omega_k}}{\bra{\omega_k} e^{-\beta U^\dagger H' U} \ket{\omega_k}}.
\end{align}
Here, the first factor on the right-hand side reproduces the standard Crooks form, whereas the second accounts for the correction appearing in the \vom scheme. This correction term may be viewed as the imprint of coherence on the fluctuation statistics: away from the commuting limit, it modifies the relative weight of forward and backward work outcomes, and it can enhance the weight of negative-work events compared with the TPM case.

\subsection{Thermodynamic uncertainty relation (TUR)}
\label{subsec:TUR}

We now develop a forward--backward thermodynamic uncertainty relation for the \vom scheme, based on the fluctuation-theorem TUR framework introduced in Refs.~\cite{Proesmans_2017, Hasegawa_2019} and extended to asymmetric forward--backward settings in Refs.~\cite{Potts_2019, Timpanaro_2024}. Let $f_k$ be any real-valued observable quantity defined on the measurement outcomes $k$, with forward mean and variance $\av{f}_\Omega := \sum_k p_k \, f_k$ and $\var_\Omega(f) := \sum_k p_k (f_k - \av{f}_\Omega)^2$.
We focus on quantities that are odd under time reversal, namely $\tilde{f}_k = -f_k$. This is the natural class of quantities for which a TUR is informative, since even quantities do not distinguish the direction of the protocol. The \vom work itself provides an example: by Eq.~\eqref{eqn:involution_outcomes}, its paired backward outcome satisfies $\tilde\omega_k=-\omega_k$.

Under this condition, we prove in Sec.~\ref{SM:TUR_derivation} of the Supplementary Material the general forward--backward \vom TUR
\begin{align} \label{eqn:TUR_Omega_general}
    \frac{\bigl(\av{f}_\Omega + \av{\tilde{f}}_{\tilde{\Omega}} \bigr)^2}{\var_\Omega (f) + \var_{\tilde{\Omega}} (\tilde{f})} \leq \mathrm{e}^{\sigma^\Omega_\mathrm{sym}} - 1,
\end{align}
where
\begin{align}
    \label{eq:Sigma_Omega^sym}
    \sigma^\Omega_\mathrm{sym} := \frac{1}{2} \bigl[D_\mathrm{KL}(p \, \Vert \, \tilde{p}) + D_\mathrm{KL}(\tilde{p} \, \Vert \, p)\bigr].
\end{align}
is the symmetrized forward--backward irreversibility measure of the VOM outcome distributions.

In particular, choosing $f_k=\omega_k$ and $\tilde f_k=-\omega_k$ yields the VOM work relation
\begin{align}
    \frac{\bigl(\av{W}_\Omega + \av{\tilde{W}}_{\tilde{\Omega}} \bigr)^2}{\var_\Omega (W) + \var_{\tilde{\Omega}} (\tilde{W})} \leq \mathrm{e}^{\sigma^\Omega_\mathrm{sym}} - 1.
\end{align}
Thus, the precision of the combined forward--backward VOM work signal is constrained by the distinguishability between the forward and backward outcome statistics.

\section{Case study: \vom work statistics in a driven qubit}
\label{subsec:case_study}

We now illustrate the \vom scheme and the associated fluctuation relations in a simple two-level example. Let us consider a qubit initially governed by the Hamiltonian $H=\frac{\Delta}{2}\sigma_z$, undergoing the coherent unitary driving $U = \ketbra{E_0}{E_+} + \ketbra{E_1}{E_-}$, where $\ket{E_\pm} = \frac{\ket{0}\pm\ket{1}}{\sqrt{2}}$, to a final Hamiltonian $H'=\frac{\Delta'}{2}\sigma_z$. This setting allows us to diagonalise the Heisenberg operator of work explicitly, construct the corresponding \vom readout, and evaluate analytically the resulting work statistics, energetic cost, and fluctuation-type relations.

\subsection{Operational implementation of the \vom scheme}

For this protocol, the Heisenberg operator of work is $\Omega = \frac{\Delta'}{2}\sigma_x - \frac{\Delta}{2}\sigma_z$. Its eigenvalues are
\begin{align}
\omega_\pm = \pm \frac{s}{2},
\label{eqn:omega_pm_E1}
\end{align}
where $s = \sqrt{(\Delta')^2 + \Delta^2}$, with corresponding normalised eigenvectors in basis $\{\ket{E_1},\ket{E_0}\}$
\begin{align}
\ket{v_\pm} = \frac{1}{N_\pm}
\begin{pmatrix}
\Delta' \\
\Delta \pm s
\end{pmatrix},
\end{align}
where $N_\pm = \sqrt{(\Delta')^2 + (\Delta \pm s)^2}$. Defining the unitary $R = \bigl[\, \ket{v_-}, \ket{v_+} \,\bigr]$, whose columns form the eigenbasis of $\Omega$ written in the energy eigenbasis, we obtain the spectral decomposition
\begin{align}
\Omega = R
\begin{pmatrix}
\omega_- & 0 \\ 0 & \omega_+
\end{pmatrix}
R^\dagger .
\end{align}

The \vom scheme is then realised by measuring $\Omega$ projectively. In the present example, this can be implemented operationally by first rotating the system into the eigenbasis of $\Omega$ through $R^\dagger$, and then performing a standard projective measurement in the energy basis. Result $s/2$ (resp.~$-s/2$) must then be attributed to the whole measurement when state $\ket{E_1}$ (resp.~$\ket{E_0}$) is obtained in the energy measurement. These two steps are enough to effectively generate the same outcome statistics as the \vom scheme. If one also wishes that the output state matches the one of a projective measurement of $\Omega$, one must finally apply the $R$ unitary.

For comparison, the TPM average work in this example is given by the average of operator $\Omega_\tpm \coloneqq \sum_{k,j} p_{j|k}(E_j' - E_k) \dyad{E_k}$, where $p_{j|k} = |\langle E_j' | U | E_k \rangle|^2$ are the TPM transition probabilities conditioned on the initial energy outcome $E_k$. Explicitly, this gives
\begin{align}
    \Omega_\tpm = -\frac{\Delta}{2}\sigma_z.
\end{align}
Comparing $\Omega$ and $\Omega_\tpm$ makes clear that the \vom and TPM schemes correspond to distinct operational notions of work already at the level of their associated observables.

\subsection{Average work in the \vom and TPM schemes}
\label{sec:energetic_cost_VOM}

We now compare the average work obtained from both schemes in this example.
We first take the initial state to be the thermal state
\begin{align}
    \tau_\beta = \frac{1}{Z} \left(e^{-\frac{\beta \Delta}{2}} \dyad{E_0} + e^{\frac{\beta \Delta}{2}} \dyad{E_1}\right),
\end{align}
with $Z = 2 \, \mathrm{cosh}\left(\frac{\beta \Delta}{2}\right)$. For this initial state, the \vom and TPM schemes assign the same average work,
\begin{align} \label{eqn:avWork_equal_E1}
    \av{W}_\Omega = \av{W}_\tpm = \frac{\Delta}{2}\tanh\left(\frac{\beta\Delta}{2}\right).
\end{align}
This agreement is expected since, as $\tau_\beta$ is diagonal in the energy basis, there is no coherence for the first TPM measurement to destroy. To highlight the differences between the two schemes in terms of average work, we must consider an input state that is coherent in the energy basis instead. Taking, for instance, 
\begin{align} \label{eq:rhocoh}
    \rho_{\mathrm{coh}} = \dyad{E_-},
\end{align} 
one obtains $\av{W}_\Omega = - \frac{\Delta'}{2}$, whereas the TPM protocol yields $\av{W}_\tpm = 0$, because the first projective energy measurement dephases $\rho_{\mathrm{coh}}$ to the maximally mixed state in the energy basis. Thus, in this example, the difference between \vom and TPM appears precisely when coherence is present in the input state.

\subsection{Jarzynski-like relation}

We next evaluate the modified Jarzynski relation for this qubit example. Since $\Omega$ has eigenvalues $\omega_\pm = \pm s/2$, the \vom exponential average takes the form
\begin{align}
\langle e^{-\beta W}\rangle_\Omega = p_- e^{\beta s/2} + p_+ e^{-\beta s/2},
\end{align}
where
\begin{align}
p_\pm = \frac{1}{2} \left[1 \pm \frac{\Delta}{s}\tanh\left(\frac{\beta\Delta}{2}\right)\right].
\label{eqn:pF_E1}
\end{align}
The correction term entering the Jarzynski-like relation is therefore
\begin{align}
\Xi_\Omega = \ln\left[\frac{\mathrm{cosh}\bigl(\frac{\beta s}{2}\bigr) -
\frac{\Delta}{s}\,\mathrm{tanh}\bigl(\frac{\beta\Delta}{2}\bigr) \,\mathrm{sinh}\bigl(\frac{\beta s}{2}\bigr)}{Z'/Z} \right],
\end{align}
where $Z' = 2\cosh\left(\frac{\beta \Delta'}{2}\right)$.

\begin{figure}[htb]
\includegraphics[width=\columnwidth]{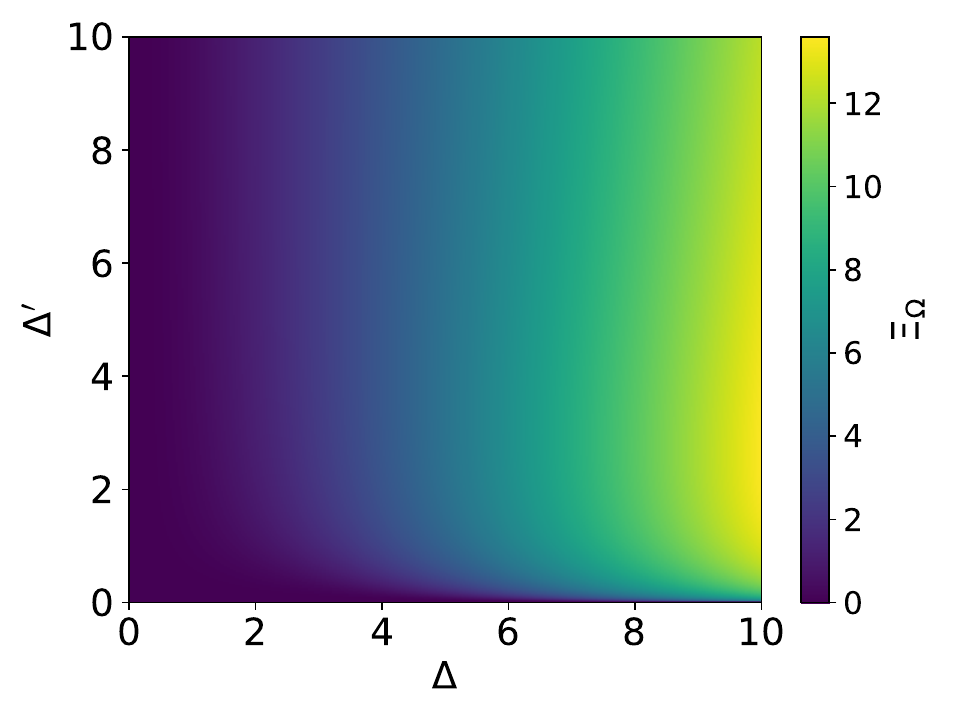}
\caption{\textbf{Jarzynski correction for the \vom scheme.} Plot of $\Xi_\Omega$ for $\Delta,\Delta' \in [0,10]$, with $\beta = 2$. The correction remains non-negative throughout the displayed parameter range, consistent with the general bound $\Xi_\Omega \geq 0$.}
\label{img:Xi_Omega}
\end{figure}

Fig.~\ref{img:Xi_Omega} shows the resulting behaviour of $\Xi_\Omega$ across the parameter range considered. As expected from the general analysis, $\Xi_\Omega$ is non-negative and quantifies the deviation of the \vom exponential average from the standard Jarzynski value.

\subsection{Crooks-like relation}

We now evaluate the corresponding Crooks-like quantity for this example. Using the forward probabilities $p_\pm$ in Eq.~\eqref{eqn:pF_E1} and the backward probabilities
\begin{align}
\tilde{p}_\pm = \frac{1}{2} \left[1 \mp \frac{\Delta'}{s}\tanh\left(\frac{\beta\Delta'}{2}\right)\right],
\label{eqn:pB_E1}
\end{align}
the Crooks-like irreversibility for this example is obtained from Eq.~\eqref{eqn:sigmaOmega_mean_KL}.

\begin{figure}[htb]
\includegraphics[width=\columnwidth]{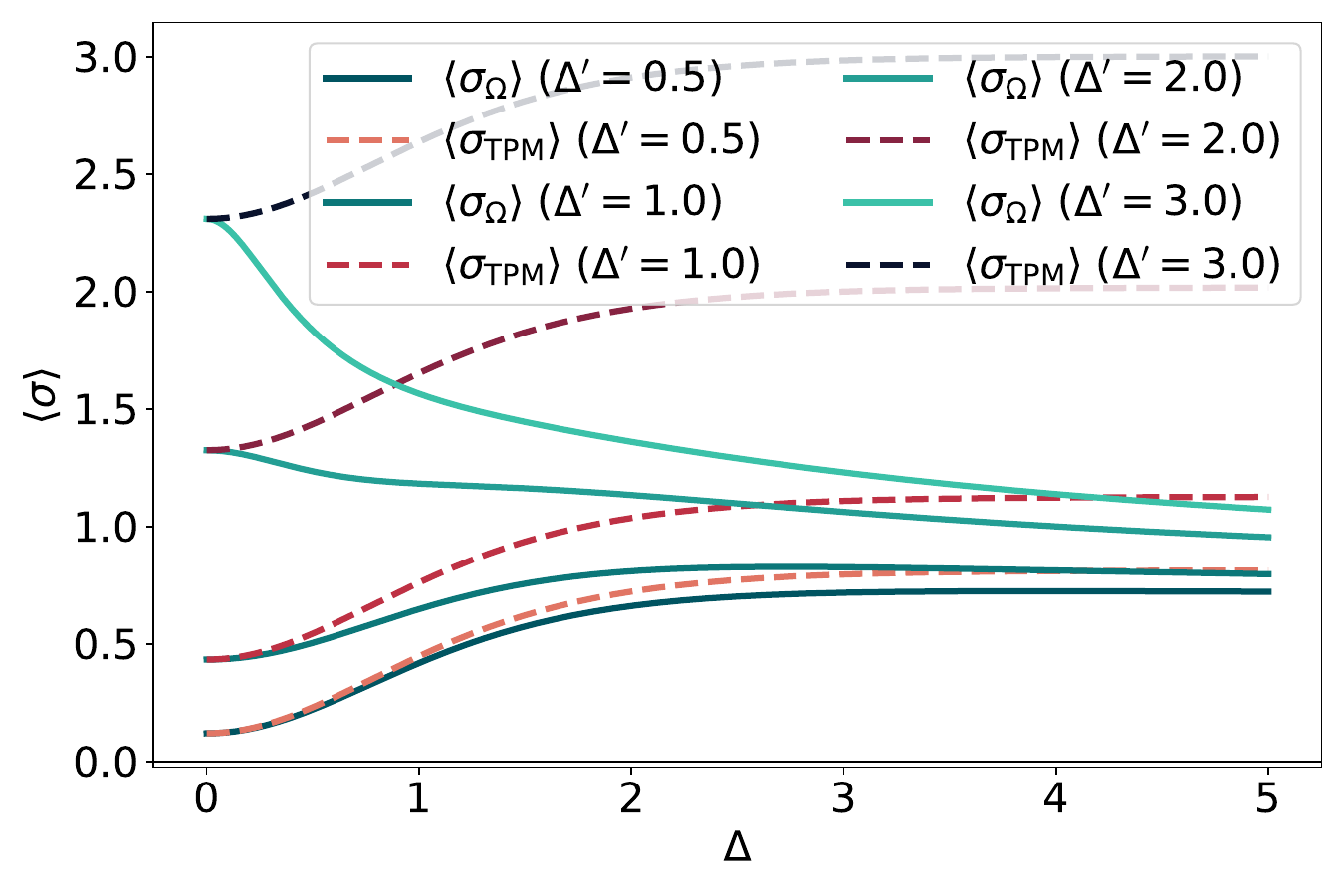}
\caption{\textbf{Crooks-like irreversibility in the \vom scheme.}
Plots of $\langle\sigma^\Omega\rangle$ as a function of $\Delta \in [0,5]$ for $\Delta' \in \{0.5,1.0,2.0,3.0\}$, with $\beta = 2$. For each fixed $\Delta'$, the \vom irreversibility measure remains below the corresponding TPM curve throughout the displayed parameter range. This reflects the fact that the two schemes quantify different operational notions of irreversibility in this example.}
\label{img:sigma_compare}
\end{figure}

Fig.~\ref{img:sigma_compare} compares $\langle\sigma^\Omega\rangle$ with the TPM expression
\begin{align}
    \av{\sigma^\tpm} = \beta \biggl[-\frac{\Delta'}{2}\tanh\left(\frac{\beta\Delta}{2}\right) - \Delta F\biggr].
\end{align}

In line with the bound derived in Sec.~\ref{subsec:Crooks-like_relation}, we find that $\av{\sigma^\Omega} \leq \av{\sigma^\tpm}$ throughout the parameter range shown in Fig.~\ref{img:sigma_compare}. The lower value of $\av{\sigma^\Omega}$ reflects the fact that the \vom scheme probes only the distinguishability retained in the outcome statistics of the work observable $\Omega$. By contrast, the TPM quantity corresponds to the full energy-trajectory description and includes the irreversibility associated with resolving initial and final energy outcomes. This is consistent with the broader operational picture developed in Sec.~\ref{sec:fluc_rel}: \vom is tailored to the average work of the original coherent process, while TPM probes a measurement-modified, energy-dephased process.

\subsection{Signal-to-noise ratio}

\begin{figure}[htb]
\includegraphics[width=0.5\textwidth]{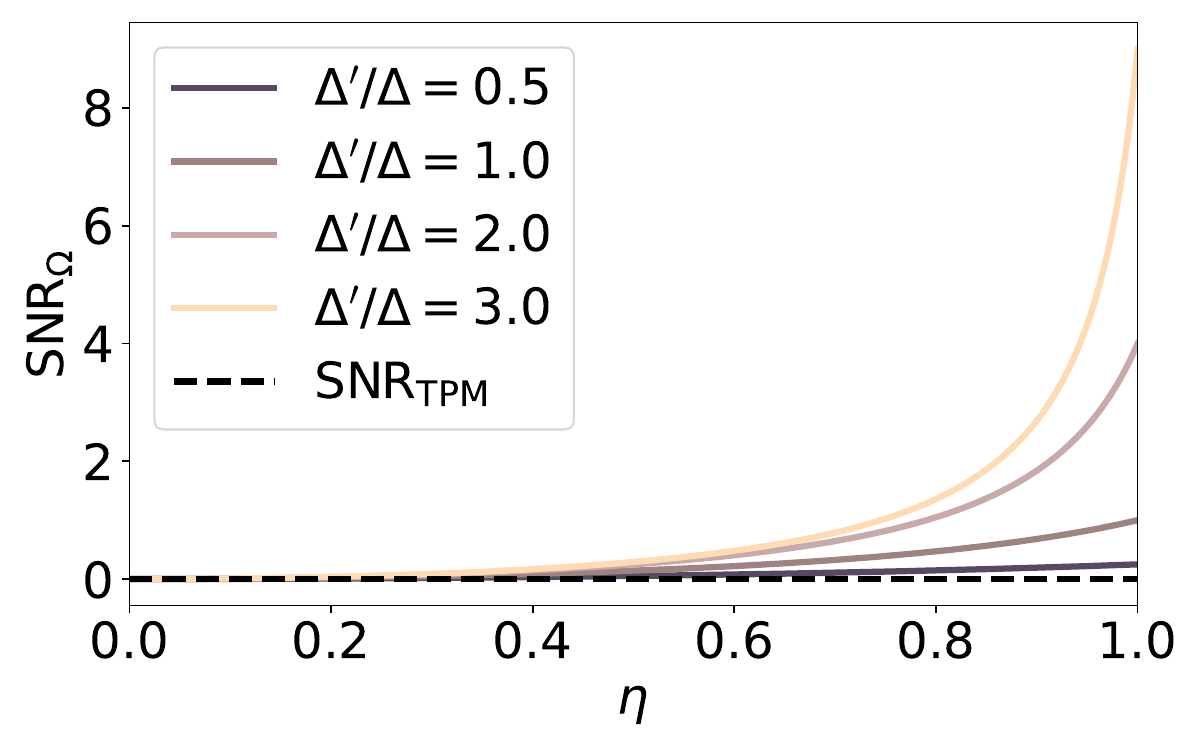}
\caption{\textbf{Coherence-enhanced signal-to-noise ratio in the \vom scheme.}
Plots of $\text{SNR}_\Omega$ and $\text{SNR}_\text{TPM}$ for $\eta \in [0,1]$ and $\Delta'/\Delta \in \{0.5,1.0,2.0,3.0\}$. The TPM signal-to-noise ratio remains identically zero throughout, whereas the \vom signal-to-noise ratio is finite for every $\eta>0$ and increases monotonically with the coherence parameter. Larger values of $\Delta'/\Delta$ lead to a stronger \vom signal, showing that even weak coherence produces a detectable work signal within the \vom scheme.} \label{img:SNR_example2} \end{figure}

For the coherent input state introduced above, the TPM and \vom schemes already differ at the level of the average work. We now quantify this difference at the level of precision by comparing the signal-to-noise ratio $\mathrm{SNR} := \frac{\av{W}^2}{\var(W)}$, which provides a metric for how the detector noise floor compares to the measured signal noise. Consider the one-parameter family of states
\begin{align}
    \rho(\eta) = (1-\eta) \frac{\id}{2} + \eta\,\rho_{\mathrm{coh}},
\end{align}
with $0\leq \eta \leq 1$, which interpolates between the maximally mixed state and the coherent pure state, and where $\rho_\text{coh}$ was defined in Eq.~\eqref{eq:rhocoh}. Since the TPM scheme dephases $\rho(\eta)$ to $\id/2$ for all $\eta$, one still has $\av{W}_\tpm = 0$, and hence $\text{SNR}_\text{TPM} = 0 \, \forall\,\eta$. For the \vom scheme, instead,
\begin{align}
    \av{W}_\Omega = \tr\bigl[\rho(\eta)\Omega\bigr] = \eta\,\tr(\rho_{\mathrm{coh}}\Omega) = -\eta\,\frac{\Delta'}{2}.
\end{align}
The second moment is then
\begin{align}
    \langle W^2\rangle_\Omega = \tr\bigl[\rho(\eta)\Omega^2\bigr] = \frac{(\Delta')^2+\Delta^2}{4}.
\end{align}
Therefore,
\begin{align}
    \var_\Omega(W) = \langle W^2\rangle_\Omega - \av{W}_\Omega^2 = \frac{\Delta^2 + (\Delta')^2(1-\eta^2)}{4},
\end{align}
and the corresponding signal-to-noise ratio is
\begin{align}
    \text{SNR}_\Omega = \frac{\eta^2(\Delta'/\Delta)^2}{1+(\Delta'/\Delta)^2(1-\eta^2)}.
\end{align}
Fig.~\ref{img:SNR_example2} shows the resulting behaviour as a function of $\eta$. While TPM predicts a vanishing signal-to-noise ratio throughout, the \vom scheme yields a finite value for every $\eta>0$. In this sense, even weak coherence produces a detectable work signal within the \vom framework, whereas TPM fails to capture it because the initial energy measurement removes the relevant coherence.

\section{Application to a four-stroke heat engine}
\label{subsec:4stroke_heat_engine}

We now turn to a concrete quantum thermal machine so as to move beyond simple one-qubit examples and assess the efficacy of the \vom scheme in a setting where coherence directly impacts performance. To this end, we revisit the four-stroke heat engine introduced in Ref.~\cite{Uzdin_2015} from the perspective of work measurement. In the regime of interest, this model exhibits coherence-assisted work extraction and thus provides a natural setting in which to compare the \vom and TPM schemes.

The working medium is a four-level system with Hamiltonian
\begin{align}
    H =& \frac{\Delta E_h}{2}\left(\dyad{E_4} -\dyad{E_1}\right) \notag\\
    & + \frac{\Delta E_c}{2}\left(\dyad{E_3} -\dyad{E_2}\right),
\end{align}
with $\Delta E_h>\Delta E_c$. During the cycle, the system alternates between thermal and unitary strokes. It first undergoes a partial cold-thermalization stroke of duration $\tau_\text{th} = 4\tau$, during which it is brought into contact with the cold bath. This is followed by a unitary driving stroke of duration $\tau_\text{dr} = \tau$, in which the baths are disconnected and an external resonant field coherently couples the transitions $\ket{E_4}\leftrightarrow\ket{E_3}$ and $\ket{E_2}\leftrightarrow\ket{E_1}$. The second half of the cycle repeats the same pattern with the hot bath: a partial hot-thermalization stroke of duration $\tau_\text{th} = 4\tau$, followed by a second unitary driving stroke identical to the first. Note that the unequal stroke durations fix the engine duty cycle, with thermal strokes longer than the coherent driving strokes so that the baths partially prepare the state before each work-extraction step.

The coherent and dissipative parts of the cycle are controlled by different parameters. The unitary work strokes are generated by the driving Hamiltonian
\begin{align}
    H_d(t)=\frac{g}{2}e^{-i\frac{(\Delta E_h-\Delta E_c)}{2}t}
    \left(\ketbra{E_4}{E_3} + \ketbra{E_2}{E_1}\right) + \mathrm{H.c.},
\end{align}
where $g$ sets the strength of the coherent drive. By contrast, the thermal strokes are described by Lindblad jump operators~\cite{Uzdin_2015}
\begin{subequations}
\begin{align}
    A_{h,\uparrow} &= \sqrt{\gamma_h} \, e^{-\beta_h \Delta E_h / 2}\ketbra{E_4}{E_1},
    \\
    A_{h,\downarrow} &= \sqrt{\gamma_h} \, \ketbra{E_1}{E_4},
\end{align}
\end{subequations}
and
\begin{subequations}
\begin{align}
    A_{c,\uparrow} &= \sqrt{\gamma_c} \, e^{-\beta_c \Delta E_c / 2}\ketbra{E_3}{E_2},
    \\
    A_{c,\downarrow} &= \sqrt{\gamma_c} \, \ketbra{E_2}{E_3}.
\end{align}
\end{subequations}
The hot-bath operators $A_{h,\uparrow}$ and $A_{h,\downarrow}$ correspond to excitation and relaxation on the $\ket{E_1}\leftrightarrow\ket{E_4}$ transition, while $A_{c,\uparrow}$ and $A_{c,\downarrow}$ do the same for $\ket{E_2}\leftrightarrow\ket{E_3}$. The Boltzmann factors are included so that the upward and downward rates have the correct thermal ratio for each bath. For simplicity, we take the two bath couplings to be equal, $\gamma_h=\gamma_c=\gamma$, so that $g$ controls the coherent driving, whereas $\gamma$ sets the strength of thermalisation. The regime we are interested in is the weak-action regime, characterised by $\gamma\tau\ll 1$ and $\gamma\ll g$. The first condition means that the thermal strokes are short enough that they do not fully relax the state within one cycle. The second means that the coherent drive remains stronger than the dissipative effects of the baths. In this regime, the working medium does not fully relax between successive strokes, and some coherence in the energy basis carries over into the next unitary step. As a result, coherence is not immediately washed out and can therefore contribute significantly to the extracted work. As shown in Ref.~\cite{Uzdin_2015}, this surviving coherence provides the dominant contribution to the extracted work.

\begin{figure}[h!]
    \centering
    \includegraphics[width=0.9\columnwidth]{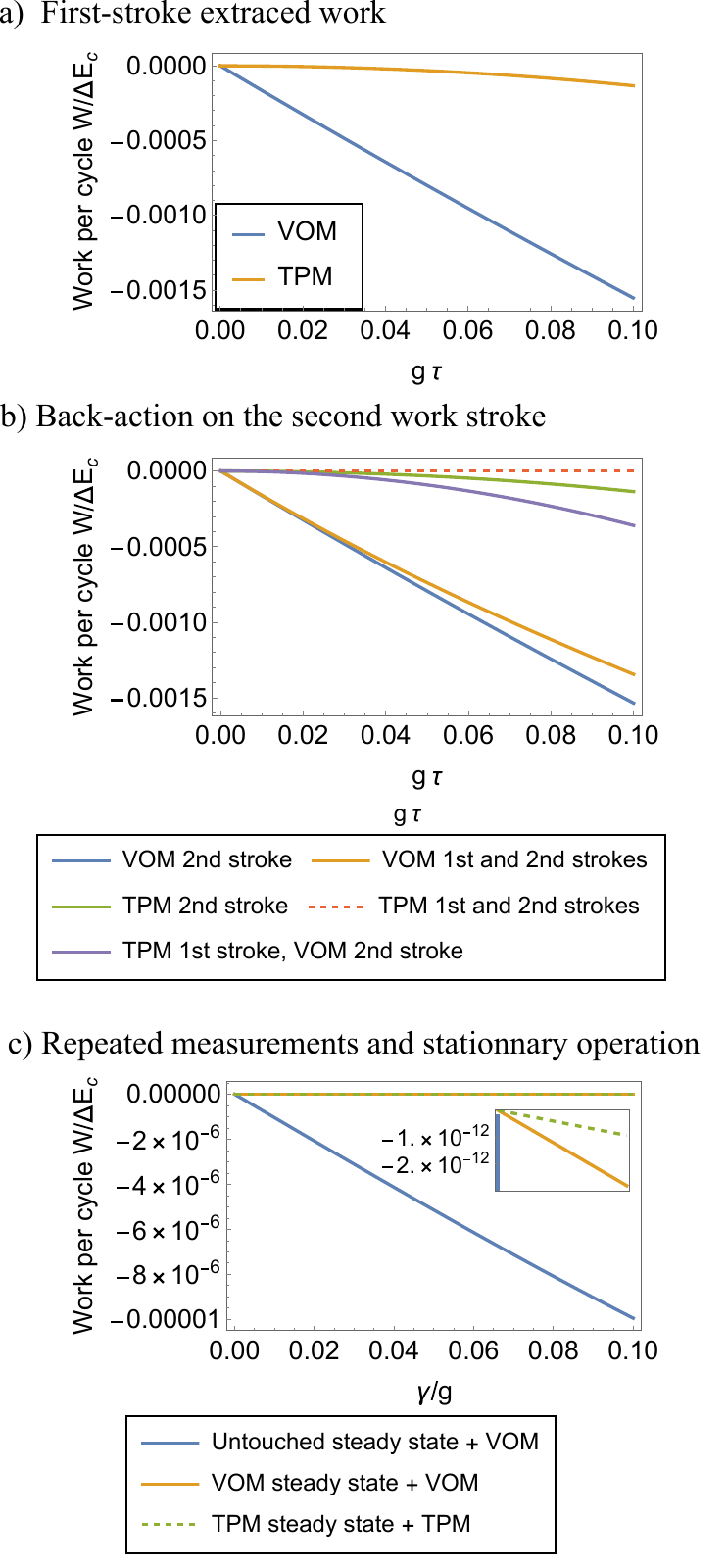}
    \caption{\textbf{Work extraction in the four-stroke engine under \vom and TPM schemes.} a) Average extracted work during the first unitary stroke as a function of the dimensionless cycle parameter $g\tau$. The \vom estimate grows linearly over the displayed range, while the TPM estimate remains much smaller, reflecting the loss of the leading coherence-assisted contribution under TPM. b) Average extracted work during the second unitary stroke for different measurement choices applied to the first and second strokes of the same cycle. Measuring the first stroke perturbs the second, but the disturbance induced by \vom remains much weaker than that induced by TPM. c) Extracted work per cycle in the stationary regime under repeated application of the schemes. Measuring with \vom at every cycle reduces the extracted work compared with the unmeasured stationary state, but the suppression remains far less severe than for repeated TPM schemes.}
    \label{fig:4stroke}
\end{figure}

For our purposes, this model is especially useful because it allows one to compare work-measurement schemes in a situation where coherence is directly tied to performance. For a coherence-assisted engine, the natural quantity to compare with is the work associated with the \textit{unmeasured} unitary stroke. This is precisely what the \vom scheme captures, since it assigns work through the spectral measurement of the corresponding Heisenberg operator of work. By contrast, TPM begins with a projective energy measurement, which dephases the state entering the stroke and therefore suppresses the coherent contribution to the extracted work. For the present engine, this distinction is especially relevant in the weak-action regime. As shown in Ref.~\cite{Uzdin_2015}, the work extracted during a short unitary stroke is linear in this pre-existing coherence, while the contribution from a fully incoherent input state appears only at higher order. As it preserves the pre-existing coherences, \vom is able to accurately capture the leading term, $\av{W}_\Omega \sim \gamma \tau$, whereas TPM removes it and yields only the weaker contribution $\av{W}_\tpm \sim (\gamma \tau)^2$. This leads to a clear operational difference: TPM underestimates the work extracted by a coherence-assisted engine, whereas \vom captures its leading contribution. The comparison is shown in Fig.~\ref{fig:4stroke}. Panel~(a) shows the average extracted work during the first unitary stroke as a function of the cycle parameter. The \vom estimate predicts an amount of work extraction that grows linearly in the range under study, while the TPM estimate remains much smaller over the same range. Panel~(b) addresses a more subtle point which is the invasiveness of the measurement scheme on the engine performance. By erasing the coherences, the TPM scheme strongly suppresses the work output of the engine rather than simply underestimating it. Although \vom is designed to diagnose the unmeasured process more faithfully than TPM, it is still a measurement and therefore still perturbs the engine. To compare the invasiveness of both schemes, we evaluate how measuring the first unitary stroke affects the state entering the second, and therefore the subsequent performance of the engine. The figure shows that this back-action remains substantially weaker for \vom than for TPM, and becomes nearly negligible in the short-cycle regime (for a discussion of the implementation of the \vom scheme with minimal back-action, see Sec.~\ref{SM:back-action} of the Supplemental Material). Finally, panel~(c) considers repeated measurements over many cycles. Here one sees that applying \vom at every unitary stroke does alter the stationary state and reduces the extracted work per cycle, but the suppression remains far less severe than under repeated application of the TPM scheme.

\section{Maxwell demon based on the \vom scheme}
\label{subsec:Maxwell_demon}

Beyond its role as a diagnostic tool, the \vom scheme also lends itself naturally to feedback-based thermodynamic protocols. The outcome $\omega_k$ of a projective measurement $\{\Pi_k\}$ of the Heisenberg work operator $\Omega$ determines the work associated with subsequently applying the unitary $U$. Once the initial state is projected onto an eigenstate $\ket{\omega_k}$, its subsequent evolution to $U\ket{\omega_k}$ satisfies  $\langle H'\rangle_{U\ket{\omega_k}} - \langle H\rangle_{\ket{\omega_k}} = \omega_k$. Thus, the work value is already fixed by the single \vom outcome. This contrasts with the TPM scheme. There, an initial energy measurement does not in general determine the work value, since the unitary $U$ can transform an energy eigenstate into a superposition of different final energy outcomes, $\ket{E_k}\to\ket{E_l}$, corresponding to distinct TPM work values. The fact that the \vom outcome already fixes the work value suggests a Maxwell-demon-type strategy in which $U$ is applied only when the measured outcome signals a favourable work-extraction event.

To illustrate this idea, we consider a qubit coupled to a hot bath at inverse temperature $\beta_h$, together with a work source able to implement a fixed unitary $U$. In contrast to the standard Maxwell-demon protocol, where the qubit is measured in the energy basis, the demon now measures the operator of work $\Omega$ associated with $U$. The cycle starts by thermalizing the qubit with the hot bath. The demon then performs a \vom, obtaining one of the outcomes $\omega_\pm$, with $\omega_-\leq 0\leq \omega_+$, and corresponding probabilities $p_\pm$. Whenever the outcome is $\omega_-$, the unitary $U$ is applied; otherwise, the qubit is left untouched. The cycle is completed by resetting the demon memory using a cold bath at inverse temperature $\beta_c > \beta_h$.

\begin{figure}[htb]
    \centering
    \includegraphics[width=\columnwidth]{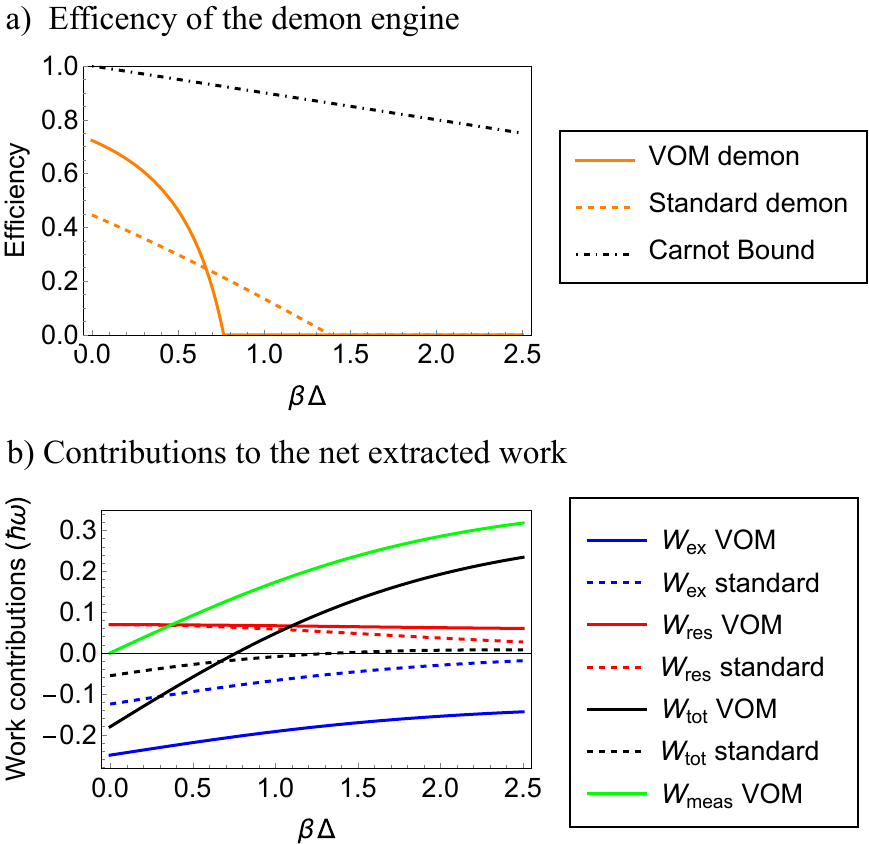}    
    \caption{\textbf{Efficiency and work-budget comparison for the \vom demon and the standard demon.} Panel (a) shows the efficiency as a function of $\beta\Delta$ for the \vom demon (solid blue), the standard demon (dashed blue), and the Carnot bound (grey), at fixed $\theta = \pi/3$. The \vom demon is slightly more efficient at small $\beta\Delta$, but its efficiency decreases more rapidly as $\beta\Delta$ increases. Panel (b) shows the corresponding contributions to the net extracted work: extracted work, reset work, and, for the \vom demon, the additional measurement work cost. The net work is shown in black for the \vom demon (solid) and the standard demon (dashed).}
    \label{fig:VOMdemon}
\end{figure}

The average work extracted during the feedback stage is then $\av{W}_{\mathrm{ex}} = p_- \omega_- \leq 0$ (where the negative sign indicates work extraction). As in a standard Maxwell demon, the memory reset carries a thermodynamic cost. For a reversible reset, this cost is
\begin{align}
    \av{W}_{\mathrm{res}} = - \beta_c^{-1} \bigl(p_+\ln p_+ + p_-\ln p_-\bigr).
\end{align}
However, unlike in the standard energy-measurement-based protocol, the \vom measurement generally does not commute with the qubit Hamiltonian and can therefore change the qubit energy already at the measurement stage. This contributes to a nonzero minimum work cost for the measurement~\cite{Latune_2025}
\begin{align}
    \av{W}_{\mathrm{meas}} = \tr\biggl[H \biggl(\sum_{\alpha = \pm} \Pi_\alpha \tau_h \Pi_\alpha - \tau_h\biggr)\biggr],
\end{align}
where $\Pi_\pm = \dyad{v_\pm}$ are the projectors onto the eigenstates of $\Omega$, and $\tau_h = e^{-\beta_h H} / Z_h$ is the thermal state prepared by the hot bath. To close the cycle, we take $H' = H$, so that no additional quench is required. The final state after the feedback step is then
\begin{align}
    \rho_f = U \Pi_- \tau_h \Pi_- U^\dagger + \Pi_+ \tau_h \Pi_+.
\end{align}
When the cycle is repeated, the qubit absorbs heat $\av{Q}_h = \tr\left[H(\tau_h - \rho_f)\right]$ from the hot bath. The total average work is therefore
\begin{align}
    \av{W}_{\mathrm{tot}} = \av{W}_{\mathrm{ex}} + \av{W}_{\mathrm{meas}} + \av{W}_{\mathrm{res}},
\end{align}
and the engine efficiency is $\eta = -\av{W}_{\mathrm{tot}} / \av{Q}_h$.

As a reference, it is useful to compare this with the standard Maxwell-demon protocol, in which the qubit is measured in the energy basis and the same unitary $U$ is applied conditionally on the outcome. The essential difference is that the standard demon bases its decision on the initial energy alone, whereas the \vom demon bases it directly on the predicted work outcome of the subsequent stroke. In this sense, the \vom demon is more closely tailored to the actual control task. As a result, it can extract more work during the feedback step than the standard demon for suitable choices of $U$. This advantage, however, comes at a price: the \vom measurement generally has an extra nonzero work cost $W_{\mathrm{meas}}$, and it can also increase the reset cost, since the latter grows with the Shannon entropy of the measurement-outcome distribution.

To make this trade-off explicit, we consider the qubit unitary $U(\theta)=e^{-i\theta \sigma_y/2}$, with $H=H'=\frac{\Delta}{2}\sigma_z$. Figure~\ref{fig:VOMdemon} compares the \vom demon and the standard demon as a function of $\beta_h \Delta$, for fixed $\theta = \pi/3$. As shown in the figure, the \vom protocol can be more efficient at high temperature, where the extra measurement cost remains relatively small, but becomes less favourable at low temperature, where the energetic cost of measuring $\Omega$ is more significant.

Panel (b) displays the three contributions that compete with each other to produce this behaviour: the work extracted during the feedback step, the memory reset cost, and, specific to the \vom protocol, the measurement work cost. The efficiency advantage of the \vom scheme is better for small rotation angles $\theta$, and vanishes for $\theta = \pi$ where $\Omega$ matches the energy basis and the \vom and standard demon measurement bases coincide.

\section*{Discussion}

This study establishes the significance of the \vom scheme as an operational measure of work in coherent quantum thermodynamics. Its relevance is most evident in regimes where coherence is an active ingredient in the dynamics, since the \vom scheme remains tied to the work associated with the original, unmeasured unitary evolution. This makes the comparison with TPM particularly instructive. TPM preserves the standard operational framework underlying the usual fluctuation relations, but, in doing so, it probes a dephased version of the process whenever coherence is present. By contrast, the \vom scheme remains sensitive to the dominant energetic contribution of coherent dynamics, but it accesses it via a projective measurement of the operator $\Omega$, which is constructed \textit{from} the unitary $U$ rather than \textit{applying} $U$ itself. In other words, the measurement is not performed while the unitary is being applied, but is rather an independent process which requires to know the unitary $U$ with good precision to be designed. Thus, the choice of work measurement scheme has direct operational consequences, as, in genuinely coherent regimes, it can alter the evaluated performance of the process itself. This should not prove surprising, given that TPM and \vom respond to different operational requirements. TPM is the natural framework if one wishes to retain the classical form of fluctuation theorems and reason upon well-defined energy variations of the system. Conversely, \vom measures the spectral outcomes of the Heisenberg work operator, and is therefore tailored to answering the question of how to assign single-shot work statistics while preserving the average energy change of the original unmeasured process. The modified fluctuation relations obtained in the \vom framework are thus appropriate for this different notion of work, and the associated correction terms can be understood as a quantitative signature of the mismatch between the TPM and \vom notions of work, which occurs away from the classical (commuting) limit.

This difference becomes especially concrete in the examples studied in this paper. In the simple few-level settings, \vom reproduces the untouched average energy change even when TPM does not. In the four-stroke engine inspired by Ref.~\cite{Uzdin_2015}, the distinction becomes even more transparent: when coherence contributes directly to the extracted work, TPM probes a dephased version of the unitary stroke, which may have significantly degraded performances -- if the engine keeps running at all without coherences. In contrast, \vom remains tied to the energetic contribution of the original coherent process and enable the coherent work extraction mechanism to occur. The point is not that \vom is completely noninvasive, no projective scheme can claim that in general, but the \vom scheme is operationally better aligned with the quantity one actually wishes to evaluate in a coherence-assisted machine. In this sense, the choice of work measurement scheme is not merely interpretational, but can decisively affect the operational assessment of performance.

This also helps to place the present results within the broader landscape of quantum work statistics. If one focuses on state-independent schemes in which work is represented by an ordinary probability distribution associated with measurement outcomes, then the coexistence of untouched average work and unmodified TPM fluctuation relations becomes impossible in coherent regimes. One may avoid this conclusion either by allowing quasiprobabilities~\cite{Allahverdyan_2014, Solinas2015, Miller2017, Lostaglio2018, Xu2018, Brodier2020, Pei_2023, Gherardini2024} or by permitting the measurement scheme to depend explicitly on the initial state~\cite{Micadei2020, Micadei2021, Micadei2024, Gherardini2021, Hovhannisyan_2024}. These alternatives are valuable in their own domains, but they address different operational goals. Quasiprobability-based approaches can encode useful information about coherent processes, although they do not correspond directly to observed work statistics when the distributions become negative or complex. State-dependent schemes, by contrast, can in principle yield proper statistics, but typically require substantial prior knowledge of the input state. From this perspective, the \vom scheme comes with both strengths and limitations. It is state-independent, produces proper measurement statistics, and reproduces the untouched average work. This comes at the cost of replacing the standard fluctuation relations with the modified relations established here. Moreover, implementing \vom requires detailed knowledge of the process through $U$, as well as sufficient control to perform the corresponding measurement of $\Omega$. In large systems, this may itself be highly nontrivial. In this sense, \vom does not remove the practical challenges of coherent work measurements, it rather shifts them, from state reconstruction or quasiprobabilistic descriptions to process characterisation and measurement control. Clarifying this balance and exploring approximate or coarse-grained versions of \vom remains an important direction for future work.

Several potential avenues for further research emerge from the present work.  One obvious next step is to compare \vom more closely with other non-TPM approaches, such as quasiprobabilistic and state-dependent schemes, to gain a clearer understanding of the operational assumptions that underpin each framework. It would also be interesting to extend the analysis to larger systems or regimes with stronger coherence, where the distinction between probing the original process and a measurement-modified one becomes clearer. Measurement back-action itself also merits further study. The appendix shows that \vom admits implementations that are optimal in precise kinematical and thermodynamic senses within the class considered there. A natural next step would be to look for explicit relations between the back-action of a scheme, its faithfulness to the average work not affected by measurement, and the size of the corrections appearing in the corresponding fluctuation relations. The examples discussed in this paper suggest that \vom is not just a formal alternative to TPM, but may also have operational significance in coherent engines and feedback-assisted thermodynamic protocols. Therefore, it would be valuable to identify realistic experimental platforms in which this distinction could be implemented and tested directly.
Extensions to open systems are also crucial for broader experimental implementation. Just as with the TPM, it is possible to consider a \vom scheme for the joint system composed of a controlled system of interest and the bath to which it is coupled. While the scheme may become challenging in this regime, investigating possible approximations in the weak system-bath coupling limit could open new avenues. Another strategy could be to replace the Heisenberg unitary evolution $U^\dagger \cdot U$ by the nonunitary evolution of the open system, yielding an observable quantifying the system energy change, which may include heat exchange with the environment.

At a more general level, the present results highlight a fundamental aspect of coherent quantum thermodynamics. When coherence plays a genuine dynamical role, it is impossible for a single operational notion of work to retain all the properties familiar from the classical limit. A selection must be made of which properties to preserve. If the aim is to preserve the TPM work variable and the standard fluctuation theorems exactly, then TPM remains the natural approach. However, if the aim is to access the energetic contribution of the original coherent dynamics without replacing it with that of a measurement-modified process, \vom provides a natural and operationally well-defined alternative.

\vspace{6mm}

\section*{Acknowledgments}
We thank G.~Landi, I.~Pikovski, N.~Yunger Halpern and M.~Williams for useful discussions. \textbf{Funding:} C.E.~acknowledges financial support from the French National Research Agency (ANR) under Grant No. ANR-22-CPJ1-0029-01. K.H.~acknowledges support from the University of Potsdam. G.R.~acknowledges financial support from the Royal Commission for the Exhibition of 1851 through a Research Fellowship, and from EPSRC through Standard Proposal Grant EP/X016218/1 (Mono-Squeeze). \textbf{Data Availability:} All codes used to produce the data are available at~\cite{RubinoVOMscheme}.

\filbreak
\renewcommand{\baselinestretch}{1.2}
\bibliography{bibliography}

\setcounter{secnumdepth}{2}
\setcounter{section}{0}
\setcounter{equation}{0}
\setcounter{figure}{0}
\setcounter{table}{0}

\appendix
\setcounter{section}{0}

\renewcommand{\thesection}{\Alph{section}}

\makeatletter
\@addtoreset{equation}{section}
\@addtoreset{figure}{section}
\@addtoreset{table}{section}
\makeatother
\renewcommand{\theequation}{\thesection\arabic{equation}}
\renewcommand{\thefigure}{\thesection\arabic{figure}}
\renewcommand{\thetable}{\thesection\arabic{table}}

\setcounter{secnumdepth}{2}
\setcounter{section}{0}
\setcounter{equation}{0}
\setcounter{figure}{0}
\setcounter{table}{0}

\appendix
\setcounter{section}{0}

\renewcommand{\thesection}{\Alph{section}}

\makeatletter
\@addtoreset{equation}{section}
\@addtoreset{figure}{section}
\@addtoreset{table}{section}
\makeatother
\renewcommand{\theequation}{\thesection\arabic{equation}}
\renewcommand{\thefigure}{\thesection\arabic{figure}}
\renewcommand{\thetable}{\thesection\arabic{table}}

\newcommand{\appsection}[1]{%
  \refstepcounter{section}%
  \section*{APPENDIX \thesection.\ \MakeUppercase{#1}}%
}

\newcommand{\appsubsection}[1]{%
  \refstepcounter{subsection}%
  \subsection*{\thesubsection.\ #1}%
}
\renewcommand{\thesubsection}{\thesection.\arabic{subsection}}

\appendix
\appsection{Can work be an observable?}
\label{app:Work_Obs}

A common objection to treating work as an observable stems from the fact that, in general, work is a property of a process rather than of a state. Consequently, its measurement is usually envisaged as a two-point, ordered-in-time process~\cite{Talkner_2007}. Indeed, in a very common approach to quantum fluctuation theorems, work is defined through two energy measurements---one before and one after the process---constituting the TPM scheme. In such a scheme, the ordering of the two measurements is part of the protocol. However, for a closed system evolution, the received work coincides with the internal energy variation, which is a state function. In that case, work only depends on the initial and final states of the evolution, or, alternatively, on the initial state and the evolution itself. Therefore, in this setting, defining work as a single operator (depending on the unitary evolution of the system) becomes physically meaningful, and its measurement---the \vom scheme, presented in the main text---is a thermodynamically consistent measurement of work.

We emphasize that, even in the closed-system setting, the distinction between the TPM and \vom schemes is fundamental when the initial Hamiltonian $H$ and the Heisenberg-evolved final Hamiltonian $\bar{H} = U^\dagger H' U$ do not commute. Namely, measuring $H$, evolving the system by $U$, and subsequently measuring $\bar{H}$ is not equivalent to performing a single measurement of the operator difference $\bar{H} - H$. In this sense, the TPM notion of fluctuating work cannot be represented by a single observable when $H$ and $\bar{H}$ do not commute \cite{Talkner_2007}. This essentially results from the way the TPM measurement scheme is intrinsically built, namely, two ordered-in-time measurements.

The same issue can also be seen at the level of the second moment. Following our definition in Eq.~\eqref{eq:Omega}, $\Omega = \bar{H} - H$, a projective measurement of $\Omega^2$ gives
\begin{equation}
    \langle\Omega^2\rangle = \langle\bar{H}^2\rangle + \langle H^2\rangle - \langle\bar{H} H\rangle - \langle H \bar{H}\rangle.
\end{equation}
The last two terms contain both operator orderings. This would be problematic if $\langle\Omega^2\rangle$ were meant to be reconstructed from two successive energy measurements, since such a procedure has a fixed temporal order. In the \vom scheme, however, this is not how the quantity is obtained. The observable $\Omega$ is measured directly, as a single Hermitian operator, and $\langle\Omega^2\rangle$ is simply the second moment of that measurement. Therefore, the appearance of both orderings does not signal an inconsistency, it simply reflects the fact that \vom is not built from a sequential measurement of $H$ and $\bar{H}$, but from the spectral measurement of their operator difference. 

All in all, the \vom scheme resolves the time-ordering difficulty by changing the operational task: work is not reconstructed from two successive measurements of generally noncommuting energy observables, but is defined through the direct spectral measurement of the single Hermitian operator $\Omega=\bar{H}-H$.

\medskip

Finally, let us discuss the connection between the closed-system and open-system scenarios. In the latter, the total system consists of our target system, now called $S$, interacting (possibly strongly) with an environment, $B$. The total system is driven only via $S$, so that the total time-dependent Hamiltonian is of the form
\begin{align*}
    H_{SB}(t) = H(t) \otimes \id_B + \id_S \otimes H_B + \underbrace{\sum_k s_k(t) \otimes b_k}_{H_\mathrm{int}(t)},
\end{align*}
where, as before, the system Hamiltonian goes from $H(t_\mathrm{i}) = H$ to $H(t_\mathrm{f}) = H'$, and the environment Hamiltonian is $H_B$. The interaction term is composed of the system-side time-dependent operators $s_k(t)$ and operators $b_k$ belonging to $B$. The total system starts in the state $\rho_{SB}$, and the unitary evolution operator for the total system is given by the time-ordered exponent
\begin{align}
    U_{SB}(t) = \overrightarrow{\exp} \Big[-i \int_{t_\mathrm{i}}^t dt' H_{SB}(t')\Big],
\end{align}
so that the state at the moment of time $t$ is $\rho_{SB}(t) = U_{SB}(t) \rho_{SB} U_{SB}(t)^\dagger$. Thus, by the standard argument \cite{Alicki_1979, Lindblad_book}, the average work done by the end of the process is
\begin{align} \nonumber
    \av{W} &= \int_{t_\mathrm{i}}^{t_\mathrm{f}} dt \tr\Big[ \rho_{SB}(t) \frac{d H_{SB}(t)}{dt} \Big]
    \\[2mm] \nonumber
    &= \tr[U_{SB}(t_\mathrm{f})\rho_{SB} U_{SB}^\dagger(t_\mathrm{f}) H_{SB}(t_\mathrm{f})] - \tr[\rho_{SB} H_{SB}(t_\mathrm{i})]
    \\[2mm] \label{totalunitarywork}
    &= \tr[\Omega_{SB} \rho_{SB}],
\end{align}
where
\begin{align}
    \Omega_{SB} = U_{SB}^\dagger(t_\mathrm{f}) H_{SB} (t_\mathrm{f}) U_{SB}(t_\mathrm{f}) - H_{SB}(t_\mathrm{i}).
\end{align}
Of course, this calculation is very well-known, and its end result \eqref{totalunitarywork} was in fact our starting point in the main text [Eq.~\eqref{requirement}]. However, we presented it here to emphasize that the operator of work---in this case $\Omega_{SB}$---captures work also in the most general open-system scenario. This means that the all our findings directly apply also there.

That said, it is important to stress that, although work is given by an operator on $SB$, it cannot be described by an operator on the level of the system $S$ whenever the action of $H_\mathrm{int}$ is nonnegligible. Another subtlety arises when approximations are made to the local dynamics of $S$ (e.g., Born--Markov \cite{bp}) or when probing capabilities are limited (e.g., locality constraints on the work measurement), in which case the inferred work may not coincide with the ``unitary work'' $\av{w}$ in Eq.~\eqref{totalunitarywork}. However, we emphasize that these limitations affect only the inferred value of work and not work itself, which remains given by Eq.~\eqref{totalunitarywork}. The interesting problem of estimating work with constrained measurement schemes is beyond the scope of this paper.

\clearpage
\onecolumngrid
\setcounter{section}{0}
\setcounter{equation}{0}
\setcounter{figure}{0}
\setcounter{table}{0}

{\centering \Large\bfseries Supplementary Material\par}

\renewcommand{\thesection}{S}

\renewcommand{\theequation}{S\arabic{equation}}
\renewcommand{\thefigure}{S\arabic{figure}}
\renewcommand{\thetable}{S\arabic{table}}

\setcounter{secnumdepth}{2}
\setcounter{section}{0}
\setcounter{subsection}{0}
\setcounter{equation}{0}
\setcounter{figure}{0}
\setcounter{table}{0}

\renewcommand{\thesection}{S\arabic{section}}
\renewcommand{\thesubsection}{\thesection.\arabic{subsection}}

\makeatletter
\renewcommand{\p@section}{}
\renewcommand{\p@subsection}{}
\makeatother

\makeatletter
\@addtoreset{equation}{section}
\@addtoreset{figure}{section}
\@addtoreset{table}{section}
\makeatother

\renewcommand{\theequation}{\thesection.\arabic{equation}}
\renewcommand{\thefigure}{S\arabic{figure}}
\renewcommand{\thetable}{S\arabic{table}}

\newcommand{\SMsection}[1]{%
  \refstepcounter{section}%
  \setcounter{subsection}{0}%
  \vspace{2.5em}%
  {\centering \bfseries S\arabic{section}. #1\par}%
  \vspace{1.5em}%
}

\newcommand{\SMsubsection}[1]{%
  \refstepcounter{subsection}%
  \vspace{2.5em}
  \subsection*{\thesubsection\ #1}%
  \nobreak\addvspace{1.5\baselineskip}
}

\SMsection{VOM scheme with minimal back-action}
\label{SM:back-action}

In this Supplementary Note, we show that the measure--evolve version of the VOM scheme has the smallest measurement back-action on the system. We demonstrate this in both kinematical and thermodynamical senses.

First of all, let us note that, in their most general form, the Kraus operators realizing a standard projective measurement of the operator $\Omega$ can be written as
\begin{align}
    K_i = \ketbra{\psi_i}{\omega_i},
\end{align}
where $\ket{\psi_i}$ are completely arbitrary. Thus, all our optimizations below will be over $\{\ket{\psi_i}\}_{i=1}^d$.

\SMsubsection{Minimizing state disturbance}
\label{app:min_state_back-action}

The measurement channel is
\begin{align}
    \Phi_\M[\rho] \coloneqq \sum_{k=1}^d \ketbra{\psi_k}{\omega_k} \rho \ketbra{\omega_k}{\psi_k}.
\end{align}
It measures the original process represented by the unitary channel
\begin{align}
    \U[\rho] \coloneqq U \rho U^\dagger.
\end{align}
Therefore, to quantify the disturbance caused by $\M$ for all input states, we will use the distance between the channels $\Phi_\M$ and $\U$. Among many possibilities, we here choose to measure it by the Bures distance \cite{Bengtsson_book_2006} between the corresponding Choi states \cite{Jamiolkowski_1972, Choi_1975}:
\begin{align} \label{choi_dist}
    \Vert \Phi_\M - \U \Vert_{\mathrm{CJ}} \coloneqq \D_{\mathrm{B}}[\J(\Phi_\M), \J(\U)].
\end{align}
This metric can \cite{DAbbruzzo_2024} and will here be called the ``Choi distance.'' Due to the unitary invariance of the Bures distance and the local unitary gauge freedom of the Choi states \cite{Paulsen_2013, Kye_2022}, the Choi distance is invariant under local basis transformations. Hence, we are free to choose any local bases for constructing the maximally entangled state. For convenience, here we choose the eigenbasis of $\Omega$, so that
\begin{align} \label{choi_state_1}
\begin{split}
    \J(\Phi_\M) &= \frac{1}{d} \sum_{k j} \Phi_\M[\ketbra{\omega_k}{\omega_j}] \otimes \ketbra{\omega_k}{\omega_j}
    \\
    &= \frac{1}{d} \sum_k \dyad{\psi_k} \otimes \dyad{\omega_k}
\end{split}
\end{align}
and
\begin{align} \label{choi_state_2}
    \J(\U) = \dyad{\J(\U)},
\end{align}
with
\begin{align} 
    \ket{\J(\U)} = \frac{1}{\sqrt{d}} \sum_k U\ket{\omega_k} \otimes \ket{\omega_k}.
\end{align}
Our goal is now to find the set of states $\ket{\psi_k}$ that minimizes $\Vert \Phi_\M - \U \Vert_{\mathrm{CJ}}$. To do so, we recall that $\D_\mathrm{B}[\rho, \sigma] = \sqrt{2\big[1 - \sqrt{\F(\rho, \sigma)} \big]}$, where $\F[\rho, \sigma] = \big(\tr \sqrt{\rho^{1/2} \sigma \rho^{1/2}} \big)^2$ is the quantum fidelity. Therefore,
\begin{align}
    \argmin_{\{\ket{\psi_k}\}} \Vert \Phi_\M - \U \Vert_{\mathrm{CJ}} = \argmax_{\{\ket{\psi_k}\}} \F \big[ \J(\U), \J(\Phi_\M) \big].
\end{align}
Finally, noting that
\begin{align}
\begin{split}
    \F \big[ \J(\U), \J(\Phi_\M) \big] &= \bra{\J(\U)} \J(\Phi_\M) \ket{\J(\U)} 
    \\
    &= \frac{1}{d^2} \sum_k |\bra{\psi_k} U \ket{\omega_k}|^2
\end{split}
\end{align}
is maximized by $\ket{\psi_k}^\mathrm{opt} = U \ket{\omega_k}$, we conclude that
\begin{align}
    \argmin_{\{\ket{\psi_k}\}} \Vert \Phi_\M - \U \Vert_{\mathrm{CJ}} = \{ U \ket{\omega_k} \}_{k=1}^d.
\end{align}
Thus, the optimized VOM scheme is the measurement of $\Omega$ implemented with the Kraus operators
\begin{align} \label{optimalKraus1}
    K^\Omega_k = U \dyad{\omega_k}.
\end{align}
Effectively, those mean: do a standard (L\"{u}ders) projective measurement of $\Omega$ at $t_\mathrm{i}$, then evolve by $U$ from $t_\mathrm{i}$ until $t_\mathrm{f}$.

\medskip

An alternative, ``time-advanced'' view on this scheme is also possible. Namely, evolve $\rho$ by $U$ from $t_\mathrm{i}$ until $t_\mathrm{f}$, then do a standard projective measurement of $U \Omega U^\dagger$ on it at $t_\mathrm{f}$. This is immediately seen by observing that
\begin{align}
    \av{W}_{\mathrm{untouched}} = \tr(\rho \Omega) = \tr(U\rho U^\dagger \, U \Omega U^\dagger).
\end{align}
As above, we note that the set of possible Kraus operators realizing this measurement is $\ketbra{\psi_k}{\omega_k} U^\dagger$, and an identical analysis leads us to the minimally disturbing Kraus operators
\begin{align} \label{optimalKraus2}
    \overline{K}_k^\Omega = U \dyad{\omega_k} U^\dagger.
\end{align}
Thus, in this case, to measure work one simply does a standard L\"{u}ders projective measurement of $\overline{\Omega} = H' - U H U^\dagger$ at the end of the process.

\medskip

Lastly, we remark that, due to the unitary invariance of the Bures distance, the VOM scheme with the Kraus operators $\dyad{\omega_i}$ is the one maximizing the fidelity between the post-measurement state and the \emph{initial state} $\rho$.

\SMsubsection{Minimizing the energetic back-action of the measurement}

When a system that starts in the state $\rho$ undergoes unitary evolution $U$, we postulate that its energy change is due to work performed by the forces driving the evolution:
\begin{align} \label{noheat}
    \av{\Delta E} = \av{W},
\end{align}
where $\av{\Delta E} = \tr(\rho \Omega)$, with $\Omega = U^\dagger H' U - H$. This generalizes the first law of thermodynamics to not-necessarily-equilibrium situations.
Importantly, this applies to the \textit{unmeasured} process.

However, when we measure the system, e.g., to estimate $W$ or $\Delta E$, we add the measurement apparatus to the mix, so the balance in Eq.~\eqref{noheat} will generally be broken.

Indeed, say, we implement a work-measuring scheme $\M = \{ K_a, w_a \}$, where $K_a$ are the Kraus operators implementing the measurement instruments and $w_a$ are the corresponding work outcomes. Then, the measured work will be
\begin{align} \label{zrt1}
    \av{W}_\M = \sum_a w_a \tr(K_a^\dagger K_a \rho).
\end{align}
While this measurement is designed to describe the $\rho \longrightarrow \rho' \coloneqq U \rho U^\dagger$ transition, it has a back-action on the state, so the actual final state is
\begin{align}
    \rho'_\M = \sum_a K_a \rho K_a^\dagger,
\end{align}
which means that the factual energy change of the system during the measuring process is
\begin{align} \label{zrt2}
    \av{\Delta E}_\M = \tr\Big[ \rho \Big( \sum_a K_a^\dagger H' K_a - H \Big)\Big].
\end{align}
This leaves us with the ``energetic back-action of the measurement''
\begin{align}
    \av{\mathfrak{B}}_\M \coloneqq \av{\Delta E}_\M - \av{W}_\M, 
\end{align}
which, after substituting the expressions for $\av{W}_\M$ and $\av{\Delta E}_\M$ from Eqs.~\eqref{zrt1} and~\eqref{zrt2} and slight rearranging, writes as
\begin{align}
     \av{\mathfrak{B}}_\M = \sum_a \tr\big[\rho K_a^\dagger (H' - W_a) K_a \big] - \tr(\rho H).
\end{align}

\medskip

For the TPM scheme, where $K_{nj} = P'_j U P_n$, with $P_n$ and $P'_j$ the eigenprojectors of, respectively, $H$ and $H'$, direct calculation shows that
\begin{align} \label{heat_for_TPM}
    \av{\mathfrak{B}}_\tpm = 0.
\end{align}

However, whenever the work measurement is such that the measured work coincides with the unmeasured work for all states; namely, when
\begin{align} \label{avenco}
    \sum_a w_a K_a^\dagger K_a = \Omega,
\end{align}
we have that
\begin{align}
    \av{\mathfrak{B}}_\M = \tr(\rho \, \theta_\M),
\end{align}
where
\begin{align}
    \theta_\M = \sum_a K_a^\dagger H' K_a - U^\dagger H' U,
\end{align}
which is generally nonzero.

\medskip

Let us now see how this plays out for the VOM scheme, where, as discussed above, $w_a \to \omega_k$ and
\begin{align}
    K_k = \ketbra{\psi_k}{\omega_k},
\end{align}
with $\ket{\psi_k}$ arbitrary. With these, we have
\begin{align} \label{ankap}
    \theta_\Omega = \sum_k  \ketbra{\omega_k}{\psi_k} H'\ketbra{\psi_k}{\omega_k} - U^\dagger H' U.
\end{align}
Here we immediately see that $\sum_k \ketbra{\omega_k}{\psi_k} H'\ketbra{\psi_k}{\omega_k}$ commutes with $\Omega$ whereas $U^\dagger H' U$ generally does not. Thus,
\begin{align}
    \theta_\Omega \neq 0 \qquad \text{whenever} \qquad [U^\dagger H' U, H] \neq 0 .
\end{align}
And therefore, $\av{\mathfrak{B}}_\Omega \neq 0$ at least for some initial states $\rho$.

To minimize the energetic back-action of the measurement universally (i.e., for all states), we minimize the Frobenius norm of $\theta_\Omega$ over $\{\ket{\psi_k}\}_k$. It is straightforward to see from Eq.~\eqref{ankap} that
\begin{align}
    \tr(\theta_\Omega^2) &= \sum_k \bra{\psi_k} H'\ket{\psi_k}^2 - 2 \sum_k \bra{\psi_k} H' \ket{\psi_k} \! \bra{\omega_k} U^\dagger H' U \ket{\omega_k} + \tr\big[ (H')^2\big] \notag
    \\
    &= \sum_k \underbrace{\big(\!\bra{\psi_k} H'\ket{\psi_k} - \bra{\omega_k} U^\dagger H' U \ket{\omega_k}\!\big)^2}_{A_k} + \underbrace{\tr\big[ (H')^2\big] - \sum_k \bra{\omega_k} U^\dagger H' U \ket{\omega_k}^2}_{B}.
    \label{zarmar}
\end{align}
Taking into account that $B \geq 0$ ($B$ is the sum of the squares of the absolute values of the off-diagonal elements of $H'$ in the $\{ U \ket{\omega_k} \}_k$ basis) and does not depend on $\ket{\psi_k}$, the minimum of Eq.~\eqref{zarmar} will be achieved when each square $A_k$ above is zero. Namely,
\begin{align}
    \ket{\psi_k}^{\mathrm{opt}} = U \ket{\omega_k}
\end{align}
minimize $\tr(\theta_\Omega^2)$. Thus, the optimal Kraus operators for the VOM scheme are
\begin{align} \label{optimalKraus3}
    K_k^{\Omega} = U \dyad{\omega_k}.
\end{align}
The minimized energetic back-action of the measurement thus reads
\begin{align}
    \av{\mathfrak{B}}_\Omega^{\mathrm{opt}} = \tr\Big[\rho \Big(\sum_k \dyad{\omega_k} U^\dagger H' U \dyad{\omega_k} - U^\dagger H' U\Big)\!\Big].
\end{align}
Note that, unless $\rho$ or $U^\dagger H' U$ commute with $\Omega$, $\av{\mathfrak{B}}_\Omega^{\mathrm{opt}} \neq 0$.

The remarkable coincidence of the Kraus operators minimizing the measurement back-action on the state [Eq.~\eqref{optimalKraus1}] and those minimizing its energetic back-action [Eq.~\eqref{optimalKraus3}] allows us to refer to the VOM measurement with Kraus operators $K_k^\Omega = U \dyad{\omega_k}$ as simply the optimal or minimal-back-action VOM scheme.

\medskip

Lastly, just as in Sec.~\ref{app:min_state_back-action}, the ``measure $\Omega$ on $\rho$, and evolve by $U$'' picture suggested by $K_k^\Omega = U \dyad{\omega_k}$ can be substituted by the equivalent, yet more intuitive, ``evolve $\rho$ by $U$, and measure $\overline{\Omega}$ on $U \rho U^\dagger$'' picture realized by $\overline{K}^\Omega_k = U \dyad{\omega_k} U^\dagger$ [Eq.~\eqref{optimalKraus2}].

\SMsubsection{Detailed energetics of the optimized VOM scheme}

We now analyze the anatomy of energy exchanges occurring during the backaction-minimized VOM protocol.

The measurement in the $\Omega$ eigenbasis changes the system energy by an amount~\cite{Jacobs_2009}
\begin{align} \label{lbmeas}
    \Delta E_\mathrm{meas} = \tr\biggl[H \biggl(\sum_k \Pi_k \rho \Pi_k - \rho\biggr)\biggr],
\end{align}
where, as always, $\Pi_k = \dyad{\omega_k}$. We note that the physical implementation of a quantum measurement comes with energetic and/or complexity overheads, such that the actual work cost may in principle significantly exceed $\Delta E_\mathrm{meas}$. In particular, it diverges in the limit in which the measurement becomes truly projective (it remains finite for a high but finite measurement efficiency)~\cite{Guryanova_2020, Latune_2025}. We will however ignore those types of overhead and focus on the measured observable-dependent part of the cost given by Eq.~\eqref{lbmeas}. One way to recover this expression is to decompose the measurement of $\Omega$ into the rotation $R^\dagger$, an energy measurement, and the inverse rotation $R$, as discussed in Sec.~\ref{subsec:VOMscheme}. The unitaries $R^\dagger$ and $R$ are reversible, closed evolutions of the system, and the corresponding energy changes can therefore be associated with work provided by the source implementing them. Adding the two contributions gives
\begin{align}
    \tr\Bigl[H\bigl(R\rho R^\dagger - \rho\bigr)\Bigr] + \tr\Bigl[H\bigl(R^\dagger\bar{\rho} R - \bar{\rho}\bigr)\Bigr] = \Delta E_\mathrm{meas},
\end{align}
where $\bar{\rho} = \sum_k \ket{E_k}\bra{E_k}R^\dagger \rho R \ket{E_k}\bra{E_k}$. The equality follows from the fact that the projection in the energy eigenbasis does not change the average energy, and using the identity $R^\dagger\bar\rho R = \sum_k \Pi_k \rho \Pi_k$.

After the measurement, in order to obtain the state with maximal fidelity to the final state of the transformation, one must apply the unitary $U$. The average work provided by the source implementing this transformation is then, taking into account that the final Hamiltonian $H'$ may differ from $H$,
\begin{align} \nonumber
    W_\mathrm{U} &= \tr\Bigl[H'U\sum_k \Pi_k\rho\Pi_kU^\dagger\Bigr]
    -\tr\Bigl[H\sum_k \Pi_k\rho\Pi_k\Bigr]
    \\
    &= \tr\Bigl[(U^\dagger H' U - H) \sum_k \Pi_k \rho \Pi_k\Bigr] = \tr(\Omega \rho) = \av{W}_{\Omega},
\end{align}
where we used the fact that the projectors $\Pi_k$ are the spectral projectors of $\Omega$.

The total energy change of the system during the protocol composed of the measurement of $\Omega$ followed by the application of $U$ is therefore $\Delta E_\mathrm{meas}+\av{W}_\Omega$. The contribution $\Delta E_\mathrm{meas}$ can thus be understood as the additional energetic cost of the VOM scheme. It is the amount of energy invested in carrying out the measurement protocol, without changing the subsequent average work exchange during the unitary $U$.

Moreover, $\Delta E_\mathrm{meas}$ is always non-negative. To see this, note that the total work $\Delta E_\mathrm{meas}+\av{W}_\Omega$ must be larger than the variation of the ergotropy of the system during the full protocol. This follows from the fact that the channel associated with the energy measurement cannot increase ergotropy, while during a unitary evolution the change in ergotropy is equal to the corresponding energy variation. Since $\av{W}_\Omega$ is equal to the energy variation of the system in the absence of the VOM, and hence to the total ergotropy variation, this bound implies that $\Delta E_\mathrm{meas}$ cannot be negative.

Finally, $\Delta E_\mathrm{meas}$ vanishes whenever the initial state $\rho$ is diagonal in the eigenbasis of $\Omega$, or whenever $[\Omega,H]=[U^\dagger H' U,H]=0$. In the first case, the average work $\av{W}_\Omega$ can still take an arbitrary value, showing that the measurement contribution $\Delta E_\mathrm{meas}$ is independent of the work assigned to the subsequent unitary stroke. The second case corresponds to the commuting regime, where the VOM and TPM schemes give the same results.

\SMsection{Derivation of the TUR bound}
\label{SM:TUR_derivation}

In this Supplementary Note, we derive Eq.~\eqref{eqn:TUR_Omega_general} in the main text. The bound obtained here is a direct application of the forward--backward thermodynamic uncertainty relation derived from fluctuation-theorem and information-theoretic arguments in Refs.~\cite{Potts_2019,Timpanaro_2024} to the VOM outcome statistics. This relation constrains the precision of observables that are odd under time reversal in terms of a forward--backward irreversibility measure.

As in the main text, we label each pair of time-reversed outcomes by a common index $k$: the forward outcome is associated with $\Pi_k$, while its backward counterpart is associated with $\tilde{\Pi}_k$ [Eq.~\eqref{eqn:proj_map}]. Their respective probabilities are $p_k$ and $\tilde{p}_k$. Consider now a quantity that is odd under this forward--backward pairing. If its forward value for outcome $k$ is $f_k$, then its value for the corresponding backward outcome is $\tilde{f}_k=-f_k$. For \vom work, this relation is guaranteed by Eq.~\eqref{eqn:OmegaB_from_OmegaF}: the forward outcome has value $f_k=\omega_k$, whereas the paired backward outcome has value $\tilde f_k=\tilde\omega_k=-\omega_k$. The backward mean can therefore be expressed using the forward-labelled values as
\begin{align}
    \av{\tilde{f}}_{\tilde{\Omega}} = \sum_k \tilde{p}_k \tilde{f}_k = - \sum_k \tilde{p}_k f_k.
\end{align}
Combining this with the forward mean $\av{f}_\Omega$, we obtain
\begin{align}
    \av{f}_\Omega + \av{\tilde{f}}_{\tilde{\Omega}} = \sum_k \bigl(p_k - \tilde{p}_k\bigr)f_k := \Delta.
\end{align}
We now introduce the symmetrised distribution
\begin{align} \label{eq:sym_distr}
    q_k:=\frac{p_k+\tilde{p}_k}{2},
\end{align}
and the function
\begin{align} \label{eq:asym_func}
    m_k:=\frac{p_k-\tilde{p}_k}{p_k+\tilde{p}_k},
\end{align}
which measures the asymmetry between forward and backward statistics. Since
\begin{align} \label{eq:p_k-tildep_k}
    p_k-\tilde{p}_k=2q_km_k,
\end{align}
we have
\begin{align} \label{eq:Delta}
    \Delta = 2 \sum_k q_k m_k f_k.
\end{align}
To express this in terms of the forward and backward variances, we define
\begin{align}
    c := \frac{\av{f}_\Omega - \av{\tilde{f}}_{\tilde{\Omega}}}{2}.
\end{align}
In this way, the mean of $f_k$ under the symmetric distribution $q$ is
\begin{align}
    \av{f}_q := \sum_k q_k f_k = \frac{1}{2} \Biggl(\sum_k p_k f_k + \sum_k \tilde{p}_k f_k\Biggr) = \frac{\av{f}_\Omega - \av{\tilde{f}}_{\tilde{\Omega}}}{2} = c.
\end{align}
Also,
\begin{align}
    \sum_k q_k m_k = \frac{1}{2} \sum_k \bigl(p_k - \tilde{p}_k \bigr) = 0
\end{align}
because both $p_k$ and $\tilde{p}_k$ are normalized probability distributions. Hence, $\sum_k q_k m_k c = 0$ and therefore we can replace $f_k$ in Eq.~\eqref{eq:Delta} by $f_k - c$:
\begin{align} \label{deltadeef}
    \frac{\Delta}{2} = \av{m_k \, (f_k-c)}_q.
\end{align}
Now, noting that, due to the Cauchy--Schwarz inequality,
\begin{align}
    \Bigl[\av{m_k \, (f_k-c)}_q\Bigr]^2 = \Bigl[\sum_k \big(m_k \sqrt{q_k}\big) \big((f_k-c) \sqrt{q_k}\big)\Bigr]^2  \leq \Bigl[\sum_k q_k m_k^2 \Bigr] \, \Bigl[ \sum_k q_k (f_k-c)^2 \Bigr],
\end{align}
we get from Eq.~\eqref{deltadeef}
\begin{align} \label{eq:CS}
    \frac{\Delta^2}{4} \leq a \,\av{(f_k-c)^2}_q,
\end{align}
where $a := \av{m^2}_q$. Furthermore, by the definition of $q_k$,
\begin{align} \label{eq:(f-c)**2}
    \langle(f_k-c)^2\rangle_q = \frac{1}{2} \sum_k p_k (f_k-c)^2 + \frac{1}{2} \sum_k \tilde{p}_k (f_k-c)^2.
\end{align}
Let us focus on the first term:
\begin{align}
    \sum_k p_k (f_k-c)^2 = \langle f^2 \rangle_\Omega - 2 c \av{f}_\Omega+ c^2 = \var_\Omega(f) + \bigl( \av{f}_\Omega - c \bigr)^2
\end{align}
where, we used the fact that $\var_\Omega(f) = \langle f^2 \rangle_\Omega - \av{f}_\Omega^2$. Analogously, for the second term in Eq.~\eqref{eq:(f-c)**2}, we find
\begin{align}
    \sum_k \tilde{p}_k (f_k-c)^2 = \var_{\tilde{\Omega}}(\tilde{f}) + \bigl(\av{\tilde{f}}_{\tilde{\Omega}} + c \bigr)^2
\end{align}
where we use the fact that $\var_{\tilde{\Omega}}(f) = \var_{\tilde{\Omega}}(\tilde{f})$ and $\av{f}_{\tilde{\Omega}} = - \av{\tilde{f}}_{\tilde{\Omega}}$. 
From the definition of $c$, we obtain that $\av{f}_\Omega - c = \av{\tilde{f}}_{\tilde{\Omega}} + c =  \frac{\av{f}_\Omega + \av{\tilde{f}}_{\tilde{\Omega}}}{2} = \frac{\Delta}{2}$, and therefore
\begin{align}
    \langle(f_k-c)^2\rangle_q = \frac{\var_\Omega(f) + \var_{\tilde{\Omega}}(\tilde{f})}{2} + \frac{\Delta^2}{4}.
\end{align}
Putting this back in Eq.~\eqref{eq:CS} we get
\begin{align}
    \frac{\Delta^2}{4} \leq a \Biggl(\frac{\var_\Omega(f) + \var_{\tilde{\Omega}}(\tilde{f})}{2} + \frac{\Delta^2}{4}\Biggr),
\end{align}
from which we obtain
\begin{align} \label{eq:interm_step}
    \frac{\bigl(\av{f}_\Omega + \av{\tilde{f}}_{\tilde{\Omega}}\bigr)^2}{\var_\Omega(f) + \var_{\tilde{\Omega}}(\tilde{f})} \leq \frac{2a}{1-a},
\end{align}
We now relate $m_k$ to the symmetrized Kullback--Leibler divergence between $p$ and $\tilde{p}$. We already defined the symmetrized forward–backward irreversibility measure $\sigma^\Omega_\mathrm{sym}$ in Eq.~\eqref{eq:Sigma_Omega^sym}. Then, by definition
\begin{align}
    D_\text{KL}(\tilde{p} \, \Vert \, p) = \sum_k \tilde{p}_k \, \ln\left(\frac{\tilde{p}_k}{p_k}\right) = - \sum_k \tilde{p}_k \, \ln\left(\frac{p_k}{\tilde{p}_k}\right),
\end{align}
and therefore
\begin{align}
    \sigma^\Omega_\mathrm{sym} = \frac{1}{2} \sum_k \bigl(p_k - \tilde{p}_k\bigr) \, \ln\left(\frac{p_k}{\tilde{p}_k}\right).
\end{align}
Due to Eq.~\eqref{eq:p_k-tildep_k} and since $p_k=q_k\bigl[1+m_k\bigr]$ and $\tilde{p}_k = q_k \bigl[1-m_k\bigr]$, we have 
\begin{align}
    \sigma^\Omega_\mathrm{sym} =  \sum_k q_k m_k \, \ln\left(\frac{1+m_k}{1-m_k}\right).
\end{align}
Introducing the function
\begin{align}
    g(x):=\sqrt{x}\,\ln\left(\frac{1+\sqrt{x}}{1-\sqrt{x}}\right), \qquad x \in [0,1),
\end{align}
we have
\begin{align}
    g(m_k^2) = m_k\,\ln\left(\frac{1+m_k}{1-m_k}\right),
\end{align}
and therefore $\sigma^\Omega_\mathrm{sym} = \sum_k q_k \, g(m_k^2)$. The function $g(x)$ is convex in $[0,1)$, therefore Jensen’s inequality gives
\begin{align} \label{eq:Sigma_ineq}
    \sigma^\Omega_\mathrm{sym} \geq g\biggl( \sum_k q_k \, m_k^2\biggr) = g(a).
\end{align}
In order to be able to compare this with Eq.~\eqref{eq:interm_step}, we want to find an expression for $a$ rather than $\sqrt{a}$. Thus, for $a\in[0,1)$, setting $x=\sqrt{a}\in[0,1)$ and using the Taylor expansions around $x=0$
\begin{subequations}
    \begin{align}
        x\ln\left(\frac{1+x}{1-x}\right)& = 2\sum_{n=0}^\infty \frac{x^{2n+2}}{2n+1} \qquad \text{ for }|x|<1,\\
        \ln\left(\frac{1+x^2}{1-x^2}\right) &= 2\sum_{n=0}^\infty \frac{x^{4n+2}}{2n+1} \qquad \text{ for }|x|<1.
    \end{align}
\end{subequations}
Subtracting term by term gives
\begin{align}
    x\ln\!\left(\frac{1+x}{1-x}\right) -\ln\!\left(\frac{1+x^2}{1-x^2}\right) = 2\sum_{n=0}^\infty \frac{x^{2n+2}(1-x^{2n})}{2n+1} \geq 0 ,
\end{align}
which leads to 
\begin{align} \label{eq:key_log_ineq}
    \sqrt{a}\,\ln\left(\frac{1+\sqrt{a}}{1-\sqrt{a}}\right) \geq \ln\left(\frac{1+a}{1-a}\right).
\end{align}
The inequality in Eq.~\eqref{eq:Sigma_ineq} then gives
\begin{align}
    \sigma^\Omega_\mathrm{sym} \geq \ln\left(\frac{1+a}{1-a}\right),
\end{align}
from which we get
\begin{align}
    \mathrm{e}^{\sigma^\Omega_\mathrm{sym}} - 1 \geq \frac{2a}{1-a}.
\end{align}
Linking this with Eq.~\eqref{eq:interm_step}, we obtain Eq. \eqref{eqn:TUR_Omega_general}, which is the final result of this derivation. Unlike a symmetric fluctuation-theorem TUR, this relation constrains the precision of the combined forward--backward signal $\av{f}_\Omega+\av{\tilde f}_{\tilde\Omega}$, rather than that of the forward mean alone.

This derivation assumes that $p_k$ and $\tilde{p}_k$ have common support, so that the logarithms are finite. If $p_k>0$ and $\tilde{p}_k=0$, or conversely, then $\sigma^\Omega_{\mathrm{sym}} = +\infty$, and the bound remains formally valid but becomes uninformative. For finite-dimensional systems at finite temperature, this condition is automatically satisfied: both Gibbs states are full rank, and every nonzero projector therefore has strictly positive probability under both the forward and backward distributions. The support caveat is relevant only in zero-temperature limits or for non-full-rank initial states.

\end{document}